# Einstein’s 1917 Static Model of the Universe: A Centennial Review

Cormac O’Raifeartaigh,[a] Michael O’Keeffe,[a] Werner Nahm[b] and Simon Mitton[c]

[a]*School of Science and Computing, Waterford Institute of Technology, Cork Road, Waterford, Ireland*

[b]*School of Theoretical Physics, Dublin Institute for Advanced Studies, 10 Burlington Road, Dublin 2, Ireland*

[c]*St Edmund’s College, University of Cambridge, Cambridge CB3 0BN, United Kingdom*

Author for correspondence: coraifeartaigh@wit.ie

## Abstract

We present a historical review of Einstein’s 1917 paper *‘Cosmological Considerations in the General Theory of Relativity’* to mark the centenary of a key work that set the foundations of modern cosmology. We find that the paper followed as a natural next step after Einstein’s development of the general theory of relativity and that the work offers many insights into his thoughts on relativity, astronomy and cosmology. Our review includes a description of the observational and theoretical background to the paper; a paragraph-by-paragraph guided tour of the work; a discussion of Einstein’s views of issues such as the relativity of inertia, the curvature of space and the cosmological constant. Particular attention is paid to little-known aspects of the paper such as Einstein’s failure to test his model against observation, his failure to consider the stability of the model and a mathematical oversight concerning his interpretation of the role of the cosmological constant. We recall the response of theorists and astronomers to Einstein’s cosmology in the context of the alternate models of the universe proposed by Willem de Sitter, Alexander Friedman and Georges Lemaître. Finally, we consider the relevance of the Einstein World in today’s ‘emergent’ cosmologies.

## 1. Introduction

There is little doubt that Einstein's 1917 paper *'Cosmological Considerations in the General Theory of Relativity'* (Einstein 1917a) constituted a key milestone in 20th century physics. As the first relativistic model of the universe, the paper, later known as 'Einstein's Static Universe' or the 'Einstein World', set the foundations of modern theoretical cosmology. In the present article, we commemorate the centenary of Einstein's 1917 paper by presenting a detailed historical analysis of the work with an emphasis on the insights it provides into Einstein's contemporaneous thoughts on relativity, astronomy and cosmology.

To be sure, a description of the basic physics of the Einstein World can be found in any standard textbook on modern cosmology (Harrison 2000 pp 355-357; Coles and Lucchin 2002 pp 26-28). However, while the historical development of theoretical cosmology from this point onwards has been described in many accounts such as (North 1965 pp 81-129; Ellis 1986; Kragh 1996 pp 7-79; Duerbeck and Seitter 2000; Nussbaumer and Bieri 2009 pp 65–110), there have been surprisingly few detailed analyses of the 1917 paper itself, and even fewer studies of the emergence of the work from the general theory of relativity in the period 1915-1917.[1] Indeed, it is probably safe to say that the paper is an example of a key scientific work that has been heavily cited but rarely analysed in detail.

The present article aims to provide a detailed review of Einstein's 1917 paper with an emphasis on the historical context of the work. Particular attention is paid to little-known aspects of this background such as: pre-relativistic models of the universe of similar geometry to the Einstein World; proposed modifications of Newton's universal law of gravity before Einstein; the problem of boundary conditions at infinity in general relativity. As regards the 1917 memoir itself, particular attention is paid to lesser-known aspects of the paper such a mathematical confusion concerning Einstein's interpretation of the cosmological constant term, Einstein's failure to test his model against observation and his failure to consider the stability of his model.

Our review is informed by primary historical resources that have become available to Einstein scholars in recent years. In particular, we refer to many letters and papers written by Einstein and his colleagues in the years 1915-1921, recently published online in English translation by Princeton University Press.[2] We also make use of the full text of Einstein's 1917 paper, shown by kind permission of the Hebrew University of Jerusalem.

[1] Some notable exceptions are (Kerzberg 1989a; Realdi and Peruzzi 2009; Weinstein 2013; Smeenk 2014).

## 2. Historical context of the Einstein World

### *(i) Biographical considerations*

Einstein's manuscript *'Kosmologische Betrachtungen zur allgemeinen Relativitätstheorie'* or *'Cosmological Considerations in the General Theory of Relativity'* (Einstein 1917a) was read to the Prussian Academy of Sciences on February 8th 1917 and published by the Academy on February 15th of that year. Thus the paper, a sizeable ten-page memoir that was to play a seminal role in 20th century cosmology, appeared only eleven months after the completion of Einstein's greatest and most substantial work, *'Die Grundlage der allgemeinen Relativitätstheorie'* or *'The Foundations of the General Theory of Relativity'* (Einstein 1916a).[3] The short interval between these two monumental papers is astonishing given that Einstein completed many other works during this period and that he suffered a breakdown in health in early 1917.[4]

Indeed, it has often been noted that the period beween autumn 1915 and spring 1917 marked a phase in Einstein's life that was extremely productive intellectually yet very difficult personally (Clark 1973 pp 190-193; Pais 1994 p18, 165; Fölsing 1997 pp 405-406). With the departure of his first wife and sons from Berlin in the summer of 1914, Einstein lived alone in a small apartment on Wittelsbacher Strasse in Berlin, working feverishly hard on the general theory of relativity and other projects and enduring a poor diet due to strained finanical circumstances and war-time food rationing. The privations of this period, possibly the most intellectually strenous of his life, led to serious health problems; from late 1916 onwards, Einstein suffered successively from liver ailments, a stomach ulcer, jaundice and general weakness. These problems were not alleviated until he was nursed back to health by Elsa Löwenthal in the summer of 1917.[5]

[2] The Collected Papers of Albert Einstein (CPAE) is an invaluable historical archive of primary sources provided by Princeton University Press in conjunction with the California Institute of Technology and the Hebrew University of Jerusalem. The collection has recently been digitized and published online with annotations and editorial comments at http://einsteinpapers.press.princeton.edu/. We make particular use of volumes 6, 7 and 8 (Kox *et al.* 1996; Janssen *et al.* 2002; Schulmann *et al.* 1998).
[3] The 'Grundlage' paper was submitted to the *Annalen der Physik* on March 20th 1916 and appeared in print on May 11th of that year (Einstein 1916a).
[4] These works include two key papers on the quantum theory of radiation (Einstein 1916b; Einstein 1916c), a paper on gravitational waves (Einstein 1916d), a paper on Hamilton's principle and general relativity (Einstein 1916e) and a popular book on relativity (Einstein 1917b).

On the other hand, it is no surprise from a scientific point of view that Einstein's first foray into cosmology should occur so soon after the completion of the general theory of relativity. After all, it was a fundamental tenet of the general theory that the geometric structure of a region of space-time is not an independent, self-determined entity, but determined by mass-energy (Einstein 1916a). Thus, considerations of the universe at large formed a natural testbed for the theory. As Einstein later remarked to the Dutch astronomer Willem de Sitter, a key motivation for his cosmological memoir was the clarification of the conceptual foundations of the general theory: *"For me, though, it was a burning question whether the relativity concept can be followed through to the finish, or whether it leads to contradictions. I am satisfied now that I was able to think the idea through to completion without encountering contradictions"* (Einstein 1917f). Indeed, it is clear from Einstein's correspondence of 1916 and early 1917 that cosmic considerations – in particular the problem of boundary conditions at infinity – were a major preoccupation in the immediate aftermath of the discovery of the covariant field equations, as will be discussed below.

Furthermore, such considerations were an important guide *throughout* the development of the general theory of relativity. As noted by analysts such as Julian Barbour (Barbour 1990), Carl Hoefer (Hoefer 1994) and Jürgen Renn (Renn 2002), Einstein's thoughts on the role of distant masses in determining the inertia of a body were an important source of inspiration in his search for the general field equations. Indeed, when describing the foundational principles of the general theory in 1918 (Einstein 1918a), he specifically cited the importance of his understanding of Mach's Principle, and of considerations of the universe at large, as will be discussed in section 2(iii). Thus there is little doubt that cosmic considerations formed an integral part of the development of the general theory of relativity, and that Einstein's quest for a consistent solution to the field equations for the case of the universe as a whole was a natural continuation of the relativity project.

[5] A description of Einstein's health problems in this period can be found in his correspondence with colleagues such as Paul Ehrenfest, Michele Besso and Hendrik Zangger (Einstein 1917c; Einstein 1917d; Einstein 1917e).

*(ii) The known universe in 1917*

In the early years of the 20th century, for most physicists and astronomers, the universe effectively comprised the Milky Way, with the density of stars decreasing drastically beyond the bounds of our galaxy (Young 1888 p511; Newcomb 1906 p33; Smith 1982 pp 55-57). Regarding the size and structure of the galaxy, a consensus had emerged that almost all of the stars lay within a round, flat disc of space, whose diameter was about eight or ten times its thickness, and whose radius was of the order of several thousand light-years (Smith 1982 p 56; Kragh 2007 pp 111-113). Studies by leading astronomers such as Jacobus Kapteyn and Hugo von Seeliger using sophisticated statistical techniques suggested that the stars were arranged in an ellipoidal distribution, with the sun near the centre (Seeliger 1898a; Kapteyn 1908; Smith 1982 p 57; Kragh 2007 pp 111-113), while the mean density of stars in the galaxy was estimated at about $10^{-23}$ g/cm$^3$ (de Sitter 1917a). Observations of globular clusters by the American astronomer Harlow Shapley were soon to extend estimates of the radius of the galaxy to well over 100,000 light-years; however, this work took place after 1917 (Shapley 1918; Smith 1982 pp 55-60).

The early years of the 20th century also saw the resurgence of an old question - whether or not the universe contained numerous galaxies of stars similar to the Milky Way. Since the time of Thomas Wright and Immanuel Kant, it had been hypothesised that the distant nebulae, cloudy entities barely discernible in the night sky with the largest telescopes, might constitute entire galaxies of stars far from our own. This 'island universes' hypothesis garnered some support during the 19th century when astronomers such as William Herschel and William Parsons observed that some nebulae displayed a spiral structure and appeared to contain stars (Smith 1982 pp 1-54). However, some doubts were cast on the hypothesis towards the end of the 19th century, with the discovery that the nebulae were clustered near the poles of the Milky Way and with the observation of an extremely bright nova in the Andromeda nebula (Smith 1982 pp 55-97).

The observational situation underwent a significant change in the 1910s with the first systematic measurements of the spectra of spiral nebulae by the American astronomer V.M. Slipher. In 1915 and 1917, Slipher published evidence that light from some of the nebulae was significantly redshifted (Slipher 1915; Slipher 1917). These observations indicated that many of the spirals were receding outwards at velocities ranging from 300 to 1100 km/s and suggested to some that they could not be gravitationally bound by the Milky Way. However, the debate could not be settled until the vast distance to the spirals was known; this data was supplied by Edwin Hubble in 1925 (Hubble 1925).

As regards theoretical cosmology, few quantitative models of the universe were proposed before 1917. One reason was the existence of several puzzles associated with the application of Newton's universal law of gravity to the universe as a whole. For example, it was not clear how a finite Newtonian universe would escape gravitational collapse, as first pointed out by the theologian Richard Bentley, a contemporary of Isaac Newton. Newton's response was to postulate a universe infinite in spatial extent in which the gravitational pull of the stars was cancelled by opposite attractions. However, he was unable to provide a satisfactory answer to Bentley's observation that such an equilibrium would be unstable.[6]

Pioneering work on non-Euclidean geometries in the late 19th century led some theoreticians to consider the possibility of a universe of non-Euclidean geometry. For example, Nikolai Lobachevsky considered the case of a universe of hyperbolic (negative) spatial curvature and noted that the lack of astronomical observations of stellar parallax set a minimum value of 4.5 light-years for the radius of curvature of such a universe (Lobachevsky 2010). On the other hand, Carl Friedrich Zöllner noted that a cosmos of spherical curvature might offer a solution to Olbers' paradox[7] and even suggested that the laws of nature might be derived from the dynamical properties of curved space (Zöllner 1872). In the United States, astronomers such as Simon Newcomb and Charles Sanders Peirce took an interest in the concept of a universe of non-Euclidean geometry (Newcomb 1898; Peirce 1891 pp 174-175), while in Ireland, the astronomer Robert Stawall Ball initiated a program of observations of stellar parallax with the aim of determining the curvature of space (Ball 1881 pp 92-93). An intriguing study of universes of non-Euclidean geometry was provided in this period by the German astronomer and theoretician Karl Schwarzschild, who calculated that astronomical observations set a lower bound of 60 and 1500 light-years for the radius of a cosmos of spherical and elliptical geometry respectively (Schwarzschild 1900). This model was developed further by the German astronomer Paul Harzer, who considered the distribution of stars and the absorption of starlight in a universe of closed geometry (Harzer 1908 pp 266-267). However, these considerations had little impact on the physics community, as most astronomers were primarily concerned with the question of the size of the Milky Way and the nature of the spiral nebulae.[8]

[6] See (Norton 1999; Kragh 2007 pp 72-74) for a discussion of the Newton-Bentley debate.
[7] The difficulty of reconciling the darkness of the night sky with a universe infinite in space and time (Kragh 2007 pp 83-86).

The end of the 19th century also saw a reconsideration of puzzles associated with Newtonian cosmology in the context of the new concepts of gravitational field and potential. Defining the gravitational potential $\Phi$ as

$$\Phi = G \int \frac{\rho\,(r)}{r} dV \tag{1}$$

where G is Newton's gravitational constant and $\rho$ is the density of matter in a volume $V$, Newton's law of gravitation could be rewritten in terms of Poisson's equation

$$\nabla^2 \Phi = 4\pi \mathrm{G}\rho \tag{2}$$

where $\nabla^2$ is the Laplacian operator. Distinguished physicists such as Carl Neumann, Hugo von Seeliger and William Thomson noted that the gravitational potential would not be defined at an infinite distance from a distribution of matter (Neumann 1896 pp 373-379; Seeliger 1985; Seeliger 1896; Thomson 1901). Neumann and Seeliger suggested independently that the problem could be solved by replacing Poisson's equation (2) with the relation

$$\nabla^2 \Phi - \lambda\Phi = 4\pi \mathrm{G}\rho \tag{3}$$

where $\lambda$ was a decay constant sufficiently small to make the modification significant only at extremely large distances.[9] A different solution to the problem was proposed in 1908 by the Swedish astronomer Carl Charlier, who considered a hierarchical or fractal structure for the universe; in this model the mean density of matter would tend to zero while the density would remain finite in every local location (Charlier 1908). This proposal was later taken up by Franz Selety, who argued that the hierarchic universe could provide a static, Newtonian cosmology alternate to Einstein's relativistic universe, as will be discussed in section 4.

---

[8] See (Kragh 2012a; Kragh 2012,b) for a review of pre-1917 models of the universe of non-Euclidean geometry and their impact.

[9] See (North 1965 pp 17-18; Norton 1999) for a review of the Neumann-Seeliger proposal.

*(iii) General relativity and the problem of boundary conditions at infinity*

In 1905, a young Einstein suggested that a ‘fixed’ interval in space or time would be measured differently by observers in uniform relative motion (Einstein 1905a). A few years later, Einstein’s erstwhile teacher Hermann Minkowski noted that, according to the theory,-the four-dimensional space-time interval

$$ds^2 = -dx^2 - dy^2 - dz^2 + c^2dt^2 \tag{4}$$

would be an invariant for such observers (Minkowski 1908). This interval is written conveniently as

$$ds^2 = \sum_{\mu,\nu=0}^{3} \eta_{\mu\nu} dx^\mu dx^\nu \tag{5}$$

where $\eta_{\mu\nu} = -1$ for $\mu = \nu = 0,1,2$, $\eta_{\mu\nu} = 1$ for $\mu = \nu = 3$ and $\eta_{\mu\nu} = 0$ for $\mu \neq \nu$. The coefficients $\eta_{\mu\nu}$ can also be written as components of the ‘Minkowski metric’

$$\begin{pmatrix} -1 & 0 & 0 & 0 \\ 0 & -1 & 0 & 0 \\ 0 & 0 & -1 & 0 \\ 0 & 0 & 0 & 1 \end{pmatrix}$$

In the general theory of relativity, the geometry of a region of space-time deviates from the ‘flat’ Minkowskian case above due to the presence of matter/energy. Thus the space-time interval $ds^2$ is written more generally as

$$ds^2 = \sum_{\mu,\nu=0}^{3} g_{\mu\nu} dx^\mu dx^\nu \tag{6}$$

where the $g_{\mu\nu}$ are gravitational potentials determined by the distribution and flux of matter/energy. In 1915, Einstein published a set of covariant field equations that specified the relation between the geometry of a region of space-time and the distribution of matter/energy within it according to

$$G_{\mu\nu} = -\kappa \left( T_{\mu\nu} - \frac{1}{2} g_{\mu\nu} T \right) \tag{7}$$

where $G_{\mu\nu}$ is a four-dimensional tensor representing the curvature of space-time (known as the Ricci curvature tensor), $T_{\mu\nu}$ is the energy-momentum tensor, $T$ is a scalar and $\kappa$ is the Einstein constant $8\pi G/c^2$ (Einstein 1915a). It was soon realised that the field equations could be alternatively written as

$$G_{\mu\nu} - \frac{1}{2} g_{\mu\nu} G = -\kappa T_{\mu\nu} \tag{8}$$

where $G$ (= $\kappa T$) is a scalar known as the Ricci curvature scalar.[10]

A description of Einstein's long path to his covariant field equations can be found in reviews such as (Norton 1984; Hoefer 1994; Janssen 2005; Janssen and Renn 2007). As noted in those references, Einstein's thoughts on Mach's Principle and the relativity of inertia played an important (if implicit) role in the development of the theory. Indeed, in his well-known *'Prinzipelles'* paper of 1918, Einstein explicitly cited three principles as fundamental in the development of the field equations (Einstein 1918a). First, the *"Principle of Relativity"* assumed that a formulation exists under which the laws of nature are invariant under arbitrary transformation: "*Nature's laws are merely statements about temporal-spatial coincidences: therefore they find their only natural expression in generally covariant equations".* Second, the *"Principle of Equivalence"* assumed that gravity and inertia are indistinguishable: *"Inertia and gravity are phenomena identical in nature. From this, and from the special theory of relativity, it follows necessarily that the symmetric "fundamental tensor" ($g_{\mu\nu}$) determines the metric properties of space, the inertial behaviour of bodies in this space, as well as the gravitational effects".* Third, *"Mach's Principle"* assumed that the metric properties of space are determined entirely by matter: *"The G-field is completely determined by the masses of the bodies. Since mass and energy are – according to the results of the special theory of relativity – the same, and since energy is formally described by the symmetric energy tensor, it follows*

[10] For purposes of clarity, we employ the nomenclature used by Einstein in the years 1915-1917. Nowadays, the Ricci curvature tensor and Ricci scalar are denoted by $R_{\mu\nu}$ and $R$ respectively.

*that the G-field is caused and determined by the energy tensor of matter"* (Einstein 1918a). Further insight into Einstein's understanding of Mach's Principle and its relevance to cosmology is offered in the same article: "*Mach's Principle (c) is a different story. The necessity to uphold it is by no means shared by all colleagues: but I myself feel it is absolutely necessary to satisfy it. With (c), according to the field equations of gravitation, there can be no G-field without matter. Obviously postulate (c), is closely connected to the space-time structure of the world as a whole, because all masses in the universe will partake in the generation of the G-field"* (Einstein 1918a).

Even before the field equations had been published in their final, covariant form, Einstein had obtained an approximate solution for the case of the motion of the planets about the sun (Einstein 1915b). In this calculation, the planetary orbits were modelled as motion around a point mass of central symmetry and it was assumed that at an infinite distance from that point, the metric tensor $g_{\mu\nu}$ would revert to the flat Minkowski space-time given by equation (4). Indeed, the orbits of the planets were calculated by means of a series of simple deviations from the Minkowski metric. The results corresponded almost exactly with the predictions of Newtonian mechanics with one exception; general relativity predicted an advance of 43" per century in the perihelion of the planet Mercury (Einstein 1915b). This prediction marked the first success of the general theory, as the anomalous behaviour of Mercury had been well-known to astronomers for some years but had remained unexplained in Newton's theory. The result was a source of great satisfaction to Einstein and a strong indicator that his new theory of gravity was on the right track.[11]

In early 1916, Karl Schwarzschild obtained the first exact solution to the general field equations, again pertaining to the case of a mass point of central symmetry (Schwarzschild 1916). Einstein was surprised and delighted by the solution, declaring in a letter to Schwarzschild in January 1916 that *"I would not have expected that the exact solution to the problem could be formulated so simply"* (Einstein 1916f). In the Schwarzschild solution, it was once again assumed that sufficiently far from a material body, the space-time metric would revert to the flat space-time of Minkowski. The imposition of such 'boundary conditions' was not unusual in field theory; however, such an approach could hardly be applied to the universe as a whole, as it raised the question of the existence a privileged frame of reference at infinity. Moreover, the assumption of a Minkowski metric an infinite distance away from matter did not chime with Einstein's understanding of Mach's Principle. These puzzles became more evident throughout 1916, as described below.

---

[11] Einstein wrote to Paul Ehrenfest that he was "beside himself with joyous excitement" at the result (Einstein 1916g) and remarked to Adriaan Fokker that the discovery gave him "palpitations of the heart" (Pais 1982 p253).

Einstein's congratulatory letter to Schwarschild casts interesting light on his view of the problem of boundary conditions at cosmic scales: *"On a small scale, the individual masses produce gravitational fields that even with the most simplifying choice of reference system reflect the character of a quite irregular small-scale distribution of matter. If I regard larger regions, as those available to us in astronomy, the Galilean reference system provides me with the analogue to the flat basic form of the earth's surface in the previous comparsion. But if I consider even larger regions, a continuation of the Galilean system providing the description of the universe in the same dimensions as on a smaller scale probably does not exist, that is, where throughout, a mass-point sufficiently removed from other masses moves uniformly in a straight line"* (Einstein 1916f). The discussion provides further insight into Einstein's view of Mach's Principle at this time: *"Ultimately, according to my theory, inertia is simply a reaction between masses, not an effect in which "space" of itself were involved, separate from the observed mass. The essence of my theory is precisely that no independent properties are attributed to space on its own"* (Einstein 1916f).

Einstein's correspondence suggests that he continued to muse on the problem of boundary conditions at infinity throughout the year 1916. For example, a letter written to his old friend Michele Besso in May 1916 contains a reference to the problem, as well as an intriguing portend of Einstein's eventual solution: *"In gravitation, I am now looking for the boundary conditions at infinity; it certainly is interesting to consider to what extent a* finite *world exists, that is, a world of naturally measured finite extension in which all inertia is truly relative"* (Einstein 1916g).

In the autumn of 1916, Einstein visited Leiden in Holland for a period of three weeks. There he spent many happy hours discussing his new theory of gravitation with his great friends Henrik Lorentz and Paul Ehrenfest.[12] Also present at these meetings was the Dutch astronomer and theorist Willem de Sitter. A number of letters and papers written shortly afterwards by de Sitter (de Sitter 1916a; de Sitter 1916b; de de Sitter 1916c; de Sitter 1916d; suggest that many of these discussions concerned the problem of boundary conditions, i.e., the difficulty of finding boundary conditions at infinity that were consistent with the Principle of Relativity and with Mach's Principle: *"In Einstein's theory all $g_{ij}$ differ from the [Minkowski] values, and*

[12] In a letter afterwards to Michele Besso, Einstein described the visit in glowing terms as "unforgettable…not only stimulating but re-invigorating" (Einstein 1916h). See also (Fölsing 1997 pp 396-398).

*they are all determined by differential equations, of which the right-hand members ($\kappa\, T_{ij}$) depend on matter. Thus matter here also appears as the source of the $g_{ij}$, i.e., of inertia. But can we say that the* whole *of the $g_{ij}$ is derived from these sources? The differential equations determine the $g_{ij}$ apart from constants of integration, or rather arbitrary functions, or boundary conditions, which can be mathematically defined by stating the values of $g_{ij}$ at infinity. Evidentially we could only say that the whole of the $g_{ij}$ is of material origin if these values at infinity were the same for all systems of co-ordinates.......... The [Minkowski] values are certainly not invariant*" (de Sitter 1916a).

In the same article, de Sitter gives evidence that, at this stage, Einstein's solution was to suggest that, at an infinite distance from gravitational sources, the components of the metric tensor $[g_{\mu\nu}]$ would reduce to degenerate values: *"Einstein has, however, pointed out a set of degenerated $g_{ij}$ which are actually invariant for all transformations in which, at infinity $x_4$ is a pure function of $x_4'$. They are:*

$$\begin{pmatrix} 0 & 0 & 0 & \infty \\ 0 & 0 & 0 & \infty \\ 0 & 0 & 0 & \infty \\ \infty & \infty & \infty & \infty^2 \end{pmatrix}$$

*.... These are then the "natural "values, and any deviation from them must be due to material sources....At very large distances from all matter the $g_{ij}$ would gradually converge towards the degenerated values"* (de Sitter 1916a).

However, de Sitter highlights a potential flaw in Einstein's proposal. Since observation of the most distant stars showed no evidence of spatial curvature, it was puzzling how the 'local' Minkowskian values of the gravitational potentials $g_{\mu\nu}$ arose from the postulated degenerate values at infinity. According to de Sitter, Einstein proposed that this effect was due to the influence of distant masses: *"Now it is certain that, in many systems of reference (i.e., in all Galilean systems) the $g_{ij}$ at large distances from all material bodies known to us actually have the [Minkowski] values. On Einstein's hypothesis, these are special values which, since they differ from [degenerate] values, must be produced by some material bodies. Consequently there must exist, at still larger distances, certain unknown masses which are the* *source of the [Minkowski] values, i.e., of all inertia* (de Sitter 1916a).[13] Yet no trace of such masses were observable by astronomy: "*We must insist on the impossibility that any of the known fixed stars or nebulae can form part of these hypothetical masses. The light even from the farthest stars and nebulae has approximately the same wavelength as light produced by terrestrial sources. ...the deviation of the $g_{ij}$ from the Galilean values ... is of the same order as here, and they must therefore be still inside the limiting envelope which separates our universe from the outer parts of space, where the $g_{ij}$ have the [degenerate] values"*. Indeed, de Sitter concludes that the hypothetical distant masses essentially play the role of absolute space in classical theory. *"If we believe in the existence of these supernatural masses, which control the whole physical universe without having ever being observed then the temptation must be very great indeed to give preference to a system of co-ordinates relatively to which they are at rest, and to distinguish it by a special name, such as "inertial system" or "ether". Formally the principle of relativity would remain true , but as a matter of fact we would have returned to the absolute space under another name" (*de Sitter 1916a).

Einstein and de Sitter debated the issue of boundary conditions at infinity in corrsepondence for some months. A review of their fascinating debate can be found in references such as (Kerzberg 1989a; Hoefer 1994; Earman 2001; Realdi and Peruzzi 2009). We note here that Einstein conceded defeat on the issue in a letter written to de Sitter on November 4th 1916: *"I am sorry for having placed too much emphasis on the boundary conditions in our discussions. This is purely a matter of taste which will never gain scientific significance. ……Now that the covariant field equations have been found, no motive remains to place such great weight on the total relativity of inertia. I can then join you in putting it this way. I always have to describe a certain portion of the universe. In this portion the $g_{\mu\nu}$ (as well as inertia) are determined by the masses present in the observed portion of space and by the $g_{\mu\nu}$ at the boundary. Which part of the inertia stems from the masses and which part from the boundary conditions depends on the choice of boundary….In practice I* must*, and in theory I* can *make do with this, and I am not at all unhappy when you reject all questions that delve further"* (Einstein 1916i). However, the closing paragraph of the same letter indicates that Einstein had not completely given up on the notion of the relativity of inertia: *"On the other hand, you must not scold me for being curious enough still to ask: Can I imagine a universe or*

[13] A similar role for distant masses is mentioned in section 2 of Einstein's 'Grundlage' paper (Einstein 1916a). The hypothesis is described in other papers and letters by de Sitter (de Sitter 1916b; de Sitter 1916c) and in contemporaneous records of the Leiden meetings (de Sitter 1916d; Peruzzi and Realdi 2011).

*the universe in such a way that inertia stems entirely from the masses and not at all from the boundary conditions? As long as I am aware that this whim does not touch the core of the theory, it is innocent; by no means do I expect you to share this curiosity"* (Einstein 1916i).

The first notice of a successful conclusion to Einstein's quest appears in another letter to de Sitter, written on 2nd February 1917: "*Presently I am writing a paper on the boundary conditions in gravitation theory. I have completely abandoned my idea on the degeneration of the* $g_{\mu\nu}$*, which you rightly disputed. I am curious to see what you will say about the rather outlandish conception I have now set my sights on*" (Einstein 1917g).[14] Another letter, written to Paul Ehrenfest two days later indicates a similar excitement and circumspection: *"I have perpetrated something ...in gravitation theory, which exposes me a bit to the danger of being committed to a madhouse. I hope there are none over there in Leyden, so that I can visit you again safely*" (Einstein 1917h). The 'outlandish conception' was the postulate of a universe of closed spatial geometry, as described below.

## 3. A guided tour of Einstein's 1917 paper

The outcome of Einstein's deliberations was the manuscript '*Kosmologische Betrachtungen zur allgemeinen Relativitätstheorie*', submitted to the Prussian Academy of Sciences on 8th February, 1917 (Einstein 1917a). The title page of the published paper is shown in figure 1. Only a fragment of Einstein's original handwritten manuscript survives, shown in figure 2 courtesy of the Hebrew University of Jerusalem. For the purposes of our analysis, we employ the standard German-English translation of the paper provided by W. Perrett and G.B. Jeffery in 1923, available online in (Einstein 1917a); we suggest that this paper be read in conjunction with this section. In a few instances, we found that the Perrett-Jeffery translation deviates slightly from the German text; such instances are highlighted in footnotes. We also note that the title of the work could have been translated as *'Cosmological Reflections on the General Theory of Relativity'* or perhaps *'Cosmological Considerations in the Context of the General Theory of Relativity'*.

Einstein's paper opens with a brief introduction which serves as an abstract. In the first paragraph of this introduction, he recalls a 'well-known' problem concerning the application of gravitational field theory to the universe at large, namely the question of the value of the gravitational potential at spatial infinity, and warns the reader that a similar problem will arise in the general theory of relativity:

> It is well known that Poisson's equation
>
> $$\nabla^2\phi = 4\pi\kappa\rho \; . \; . \; . \; . \; \text{(E1)}$$ [15]
>
> in combination with the equations of motion of a material point is not as yet a perfect substitute for Newton's theory of action at a distance. There is still to be taken into account the condition that at spatial infinity the potential $\phi$ tends towards a fixed limiting value. There is an analogous state of things in the theory of gravitation in general relativity. Here too, we must supplement the differential equations by limiting conditions at spatial infinity if we really have to regard the universe as being of infinite spatial extent.

[14] It is sometimes stated that the first notice of Einstein's solution to the problem of boundary conditions at infinity appears in a letter to Michele Besso dated December 2016 (Speziali 1955 p58; Kerzberg 1989a; Realdi and Peruzzi 2012). It is now known that this letter was written in March 1917 (Einstein 1917d).

In the second paragraph of his introductory section, Einstein recalls that when he applied general relativity to the motion of the planets, a reference frame was chosen in which the gravitational potentials became constant at spatial infinity. He warns that the same approach may not be applicable for the case of the universe at large and announces that the current paper will present his reflections on "this fundamentally important question":

> In my treatment of the planetary problem I chose these limiting conditions in the form of the following assumption: it is possible to select a system of reference so that at spatial infinity all the gravitational potentials $g_{\mu\nu}$ become constant. But it is by no means evident *a priori* that we may lay down the same limiting conditions when we wish to take larger portions of the physical universe into consideration. In the following pages, the reflexions will be given which, up to the present, I have made on this fundamentally important question.

**§1. The Newtonian Theory**

In the first section of his cosmological memoir, Einstein presents a detailed analysis of the shortcomings of Newtonian mechanics when applied to the universe as a whole and proposes a simple, but radical modification of Newton's law of gravity as solution. This section has the dual purpose of introducing the reader to the concept of a mathematical model of the universe, and of introducing a modification of Newtonian gravitation that will set the stage for a necessary modification of the relativistic field equations.

In the first paragraph of section §1, Einstein recalls that, in order to avoid the hypothesis of an infinitely large gravitational force acting on a material particle, one is led to the hypothesis of a finite island of stars in the infinite ocean of space:

> It is well known that Newton's limiting condition of the constant limit for $\phi$ at spatial infinity leads to the view that the density of matter becomes zero at infinity. For we

[15] We have relabelled Einstein's equations (1-15) as (E1-E15) in order to avoid confusion with our own article. In the original German text, Einstein uses the symbol $\Delta$ for the Laplacian operator $\nabla^2$ (figure 1a). The constant $\kappa$ in equation (E1) denotes the gravitational constant G.

> imagine that there may be a place in universal space (central point)[16] around about which the gravitational field of matter, viewed on a large scale, possesses spherical symmetry. It then follows from Poisson's equation that, in order that $\phi$ may tend to a limit at infinity, the mean density $\rho$ must decrease toward zero more rapidly than $1/r^2$ as the distance $r$ from the centre increases. In this sense, therefore, the universe according to Newton is finite, although it may possess an infinitely great total mass.

This problem associated with Newtonian models of the cosmos was discussed in section 2(ii) above. A year later, Einstein restated the problem more simply as: *"The stellar universe ought to be a finite island in the infinite ocean of space"* (Einstein 1918b, p123). Einstein then invokes a statistical argument to highlight a problem associated with such a model of the universe, namely a process of gradual evaporation of the stars:

> From this it follows in the first place that the radiation emitted by the heavenly bodies will, in part, leave the Newtonian system of the universe, passing radially outwards, to become ineffective and lost in the infinite. May not entire heavenly bodies do likewise? It is hardly possible to give a negative answer to this question. For it follows from the assumption of a finite limit for $\phi$ at spatial infinity that a heavenly body with finite kinetic energy is able to reach spatial infinity by overcoming the Newtonian forces of attraction. By statistical mechanics this case must occur from time to time, as long as the total energy of the stellar system – transferred to one single star – is great enough to send that star on its journey to infinity, whence it can never return.

One solution is to postulate a very large value for the gravitational potential at infinity. However, such a postulate is at odds with astronomical observation:

> We might try to avoid this peculiar difficulty by assuming a very high value for the limiting potential at infinity. That would be a possible way, if the gravitational potential were not itself necessarily conditioned by the heavenly bodies. The truth is that we are compelled to regard the occurrence of any great differences of potential of the gravitational field as contradicting the facts. These differences must really be of so low an order of magnitude that the stellar velocities generated by them do not exceed the velocities actually observed.

This point reflects an observational argument made by de Sitter in November 1916 (de Sitter 1916a) as discussed in section 2(iii). Einstein then notes that the model is also problematic if one applies Boltzmann statistics to the stars:

> If we apply Boltzmann's law of distribution for gas molecules to the stars, by comparing the stellar system with a gas in thermal equilibrium, we find that the Newtonian stellar system cannot exist at all. For there is a finite ratio of densities corresponding to the finite difference of potential between the centre and spatial infinity. A vanishing of the density at infinity thus implies a vanishing of the density of the centre.

[16] The words "central point" or *"Mittelpunkt*" are missing in the Perrett-Jeffery translation.

Einstein then suggests a solution to the problem, namely a simple modification of the Newtonian law of gravity. He notes in advance that the proposed modification should not to be taken too seriously but should be considered as a 'foil' for the relativistic case:

> It seems hardly possible to surmount these difficulties on the basis of the Newtonian theory. We may ask ourselves the question whether they can be removed by a modification of the Newtonian theory. First of all, we will indicate a method that does not in itself claim to be taken seriously; it merely serves as a foil for what is to follow.

The analysis to follow is worked out from first principles. Einstein notes that his proposed modification of Newton's law of gravitation allows for an infinite space filled with a uniform distribution of matter, unaware [17] that a similar modification was earlier proposed by Hugo von Seeliger (Seeliger 1895; Seeliger 1896) and by Carl Neumann (Neumann 1896), as discussed in section 2(ii):

> In place of Poisson's equation we write
>
> $$\nabla^2\phi - \lambda\phi = 4\pi\kappa\rho \quad . \quad . \quad . \quad . \quad . \qquad \text{(E2)}$$
>
> where λ denotes a universal constant. If $\rho_0$ be the uniform density of a distribution of mass, then
>
> $$\phi = -\frac{4\pi\kappa}{\lambda}\rho_0 \quad . \quad . \qquad . \qquad \text{(E3)}$$
>
> is a solution of equation (2). This solution would correspond to the case in which the matter of the fixed stars was distributed uniformly through space, if the density $\rho_0$ is equal to the actual mean density of the matter in the universe. The solution then corresponds to an infinite extension of the central space, filled uniformly with matter.

Einstein points out that the new solution reduces to the old in the neighbourhood of stars:

> If, without making any change in the mean density, we imagine matter to be non-uniformly distributed locally, there will be, over and above the $\phi$ with the constant value of equation (3), an additional $\phi$, which in the neighbourhood of denser masses will so much the more resemble the Newtonian field as $\lambda\phi$ is smaller in comparison with $4\pi\kappa\rho$.

Thus, a simple modification of Newton's law of gravitation has overcome the problem of the equilibrium of matter in an infinite, static universe:

> A universe so constituted would have, with respect to its gravitational field, no centre. A decrease of density in spatial infinity would not have to be assumed, but both the mean potential and the mean density would remain constant to infinity. The conflict with statistical mechanics which we found in the case of Newtonian theory is not repeated. With a definite but extremely small density, matter is in equilibrium, without any internal material forces (pressures) being required to maintain equilibrium.

---

[17] It is sometimes assumed (Norton 1999; Earman 2001) that Einstein knew of Seeliger's modification of Newton's universal law of gravitation when writing his cosmological memoir. However, the first reference in Einstein's writings to Seeliger's work is found in a letter to Rudolf Förster in November 1917 (Einstein 1917i). From this point onwards, Einstein cited Seeliger scrupulously (Einstein 1918b p123; Einstein 1919a; Einstein 1931; Einstein 1933). In his 1919 paper, Einstein remarked that he was unaware of Seeliger's work when writing his 1917 memoir (Einstein 1919a). See also (Kragh 2015 p63).

**§2. The boundary conditions according to the general theory of relativity**

In the second section of his paper, Einstein gives a brief history of the problem of formulating boundary conditions for spatial infinity in relativistic cosmology. The discussion is preceded by an intriguing reference to his 'rough and winding road' to a solution, and an advance warning that a modification of the field equations, analogous to the modification of Newtonian mechanics of the preceding section, will be required:

> In the present paragraph I shall conduct the reader over the road that I have myself travelled, rather a rough and winding road, because otherwise I cannot hope he will take much interest in the result at the end of the journey. The conclusion I shall arrive at is that the field equations of gravitation which I have championed hitherto still need a slight modification, so that on the basis of general relativity those fundamental difficulties may be avoided which have been set forth in §1 as confronting the Newtonian theory. This modification corresponds perfectly to the transition from Poisson's equation (1) to equation (2) of §1.

We note that Einstein's claim that the modification of the field equations to come *"corresponds perfectly"* to the modification of Newtonian gravity of section **§**1 is not quite accurate, as will be discussed in section 4. He also gives an advance preview of his solution to the problem of boundary conditions, namely the postulate of *"a self-contained continuum of finite spatial volume":*

> We finally infer that boundary conditions in spatial infinity fall away altogether, because the universal continuum in respect of its spatial dimensions is to be viewed as a self-contained continuum of finite spatial (three-dimensional) volume.

Einstein then describes his initial approach to the problem of boundary conditions at infinity. The starting point is a clear statement of his understanding of the relativity of inertia:

> The opinion which I entertained until recently, as to the limiting conditions to be laid down in spatial infinity, took its stand on the following considerations. In a consistent theory of relativity, there can be no inertia *relative to "space",* but only an inertia of masses *relative to one another*. If, therefore, I have a mass at a sufficient distance from all other masses in the universe, its inertia must fall to zero. We will try to formulate this condition mathematically.

He then derives the components of the energy-momentum tensor. Comparing these components to an element of a space-time that is assumed to be isotropic, he finds that the postulate of the relativity of inertia implies a degeneration of the gravitational potentials at infinity:

> According to the general theory of relativity, the negative momentum is given by the first three components, the energy by the last component of the covariant tensor multiplied by the $\sqrt{-g}$

$$m\sqrt{-g}\ g_{\mu\alpha}\frac{dx_\alpha}{ds}\ .\ .\ .\ .\ (\text{E4})$$

where, as always, we set

$$ds^2 = g_{\mu\nu}dx_\mu dx_\nu\ .\ .\ .\ .\ (\text{E5})$$

In the particularly perspicuous case of the possibility of choosing the system of co-ordinates so that the gravitational field at every point is spatially isotropic, we have more simply

$$ds^2 = -\,\mathrm{A}(dx_1^2 + dx_2^2 + dx_3^2) + \mathrm{B}dx_4^2$$

If, moreover, at the same time

$$\sqrt{-g} = 1 = \sqrt{A^3B}$$

we obtain from (4), to a first approximation for small velocities,

$$m\frac{A}{\sqrt{B}}\frac{dx_1}{dx_4}, m\frac{A}{\sqrt{B}}\frac{dx_2}{dx_4}, m\frac{A}{\sqrt{B}}\frac{dx_3}{dx_4}$$

for the components of momentum, and for the energy (in the static case)

$$m\sqrt{B}$$

From the expressions for the momentum, it follows that $m\frac{A}{\sqrt{B}}$ plays the part of the rest mass. As $m$ is a constant peculiar to the point of mass, independently of its position, this expression, if we retain the condition $\sqrt{-g} = 1$ at spatial infinity, can vanish only when A diminishes to zero, while B increases to infinity. It seems therefore that such a degeneration of the coefficients $g_{\mu\nu}$ is required by the postulate of the relativity of all inertia. This requirement implies that the potential energy $m\sqrt{B}$ becomes infinitely great at infinity.[18]

Einstein notes that his analysis appears to overcome the 'evaporation' problem of the Newtonian case. He also notes that the result stands independent of the simplifying assumptions employed:

> Thus a point of mass can never leave the system; and a more detailed investigation shows that the same thing applies to light-rays. A system of the universe with such behaviour of the gravitational potentials at infinity would not therefore run the risk of wasting away which was mooted just now in connection with the Newtonian theory.
> I wish to point out that the simplifying assumptions as to the gravitational potentials on which this reasoning is based, have been introduced merely for the sake of lucidity. It is possible to find general formulations for the behaviour of the $g_{\mu\nu}$ at infinity which express the essentials of the question without further restrictive assumptions.

Einstein then turns to the specific problem of the gravitational field of the stellar system. Assuming a static system of central symmetry, he announces that it proved impossible to reconcile the system with the postulate of degenerate boundary conditions, a point previously made by de Sitter:

[18] The condition $\sqrt{-g} = 1$ is used to simplify the analysis, in a manner similar to that of the *'Grundlage'* paper of 1916 (Einstein 1916a). It was later realised that the imposition of such unimodular co-ordinates can lead to a slightly different version of the field equations (Janssen and Renn 2004).

> At this stage, with the kind assistance of the mathematician J. Grommer,[19] I investigated centrally symmetrical, static gravitational fields, degenerating at infinity in the way mentioned. The gravitational potentials $g_{\mu\nu}$ were applied, and from them, the energy-tensor $T_{\mu\nu}$ of matter was calculated on the basis of the field equations of gravitation. But here it proved that for the system of the fixed stars no boundary conditions of the kind can come into the question at all, as was also rightly emphasized by the astronomer de Sitter recently.[20]

Einstein shows explicitly how the problem arises, assuming that stellar velocities are much smaller than the speed of light:

> For the contravariant energy-tensor $T^{\mu\nu}$ of ponderable matter is given by
>
> $$T^{\mu\nu} = \rho \frac{dx_\mu}{ds}\frac{dx_\nu}{ds},$$
>
> where $\rho$ is the density of matter in natural measure. With an appropriate choice of the system of co-ordinates, the stellar velocities are very small in comparison with that of light. We may therefore substitute $\sqrt{g_{44}}dx_4$ for $ds$. This shows us that all components of the $T^{\mu\nu}$ must be very small in comparison with the last component $T^{44}$. But it was quite impossible to reconcile this condition with the chosen boundary conditions.

Einstein notes that the result is not that surprising from a physical point of view. As was argued in the Newtonian case, astronomical observations suggest that the gravitational potential of distant stars cannot be much greater than that on earth:

> In retrospect this result does not appear astonishing. The fact of the small velocities of the stars allows the conclusion that wherever there are fixed stars, the gravitational potential (in our case $\sqrt{B}$) can never be much greater than here on earth. This follows from statistical reasoning, exactly as in the case of the Newtonian theory. At any rate, our calculations have convinced me that such conditions of degeneration for the $g_{\mu\nu}$ in spatial infinity may not be postulated.

Einstein now discusses two other approaches to the problem, the assumption of Minkowskian metric at infinity or the abandonment of a general solution (see also figure 2):

> After the failure of this attempt, two possibilities next present themselves.
> (a)We may require, as in the problem of the planets, that, with a suitable choice of the system of reference, the $g_{\mu\nu}$ in spatial infinity approximate to the values
>
> $$\begin{matrix} -1 & 0 & 0 & 0 \\ 0 & -1 & 0 & 0 \\ 0 & 0 & -1 & 0 \\ 0 & 0 & 0 & +1 \end{matrix}$$
>
> (b) We may refrain entirely from laying down boundary conditions for spatial infinity claiming general validity; but at the spatial limit of the domain under consideration we have to give the $g_{\mu\nu}$ separately in each individual case, as hitherto we were accustomed to give the initial conditions for time separately.

---

[19] The Jewish mathematician Jakob Grommer collaborated with Einstein on a number of works in these years (Pais 1982) p487.

[20] As discussed in section 2(iii), de Sitter highlighted the problem of reconciling degenerate values for the gravitational potentials with astronomical observation (de Sitter 1916a; de Sitter 1916b).

The second possibility amounts to abandoning the search for boundary conditions, as suggested by de Sitter in their correspondence (see section 2(iii)), but this option is not attractive to Einstein:

> The possibility (b) holds out no hope of solving the problem but amounts to giving it up. This is an incontestable position, which was taken up at the present time by de Sitter.[21] But I must confess that such a complete resignation in this fundamental question is for me a difficult thing. I should not make my mind up to it until every effort to make headway toward a satisfactory view had proved to be in vain.

On the other hand, option (a) pre-supposes a preferred frame of reference, in contradiction with basic principles of relativity. Further objections are that option (a) is not compatible with the relativity of inertia and that it does not overcome the statistical problem articulated for the Newtonian case:

> Possibility (a) is unsatisfactory in more ways than one. In the first place, those boundary conditions pre-suppose a definite choice of reference, which is contrary to the spirit of the relativity principle. Secondly, if we adopt this view, we fail to comply with the requirement of the relativity of inertia. For the inertia of a material point of mass m (in natural measure) depends upon the $g_{\mu\nu}$; but these differ but little from their postulated values, as given above, for spatial infinity. Thus inertia would indeed be influenced, but would not be conditioned by matter (present in finite space). If only one single point of mass were present, according to this view, it would possess inertia, and in fact an inertia almost as great as when it is surrounded by the other masses of the actual universe. Finally, those statistical objections must be raised against this view which were mentioned in respect of the Newtonian theory.

Einstein concludes that he has not succeeded in formulating boundary conditions for spatial infinity. Instead, he announces a new way out, the postulate of a universe of closed spatial geometry:

> From what has now been said it will be seen that I have not succeeded in formulating boundary conditions for spatial infinity. Nevertheless, there is still a possible way out, without resigning as suggested under (b). For if it were possible to regard the universe as a continuum, which is finite (closed) with respect to its spatial dimensions, we should have no need at all of such boundary conditions.

Einstein also warns the reader that his 'spatially finite' solution will come at a price, namely a modification of the field equations:

> We shall proceed to show that both the general postulate of relativity and the fact of the small stellar velocities are compatible with the hypothesis of a spatially finite universe; though certainly, in order to carry through this idea, we need a generalizing modification of the field equations of gravitation.

[21] A footnote at this point makes specific reference to (de Sitter 1916b).

## §3. The spatially closed universe with a uniform distribution of matter[22]

Two fundamental assumptions of Einstein's model are reflected in the title of this section; assuming a model of the cosmos that is spatially closed, he also assumes that, on the very largest scales, the distribution of matter is uniform and thus the curvature of space is constant:

> According to the general theory of relativity, the metrical character (curvature) of the four-dimensional space-time continuum is defined at every point by the matter at that point and the state of that matter. Therefore, on account of the lack of uniformity in the distribution of matter, the metrical structure of this continuum must necessarily be extremely complicated. But if we are concerned with the structure only on a large scale, we may represent matter to ourselves as being uniformly distributed over enormous spaces, so that its density of distribution is a variable function which varies extremely slowly. Thus our procedure will somewhat resemble that of the geodesists who, by means of an ellipsoid, approximate to the shape of the earth's surface, which on a small scale is extremely complicated.

A third assumption is that there exists a reference frame in which matter is at rest: this assumption is based on what Einstein terms the "most important fact we draw from experience as to the distribution of matter", the low velocities of the stars:

> The most important fact that we draw from experience as to the distribution of matter is that the relative velocities of the stars are very small as compared with the velocity of light. So I think that for the present we may base our reasoning upon the following approximate assumption. There is a system of reference relatively to which matter may be looked upon as being permanently at rest.

Einstein then embarks on a simple analysis in which he derives values for the components of the field equation tensors. First he derives the energy-momentum tensor, assuming a uniform density of matter:

> With respect to this system, therefore, the contravariant energy tensor $T^{\mu\nu}$ of matter is, by reason of (5), of the simple form
>
> $$\left.\begin{matrix} 0 & 0 & 0 & 0 \\ 0 & 0 & 0 & 0 \\ 0 & 0 & 0 & 0 \\ 0 & 0 & 0 & \rho \end{matrix}\right\} \quad . \quad . \quad .(E6)$$
>
> The scalar $\rho$ of the (mean) density of distribution may be *a priori* a function of the space co-ordinates. But if we assume the universe to be spatially closed, we are prompted to the hypothesis that $\rho$ is to be independent of locality. On this hypothesis we base the following considerations.

Next he turns to the gravitational potentials, starting with the assumed independence of the time co-ordinate:

---

[22] The expression *"Die räumlich geschlossene Welt"* is translated by Perrett and Jeffrey as "the spatially finite universe". We consider "the spatially closed universe" a more accurate translation.

As concerns the gravitational field, it follows from the equation of motion of the material point

$$\frac{d^2x_\nu}{ds^2} + \{\alpha\beta, \nu\}\frac{dx_\alpha}{ds}\frac{dx_\beta}{ds} = 0$$

that a material point in a static gravitational field can remain at rest only when $g_{44}$ is independent of locality. Since, further, we presuppose independence of the time co-ordinate $x_4$ for all magnitudes, we may demand for the required solution that, for all $x_\nu$,

$$g_{44} = 1 \quad . \quad . \quad . \quad . \qquad \text{(E7)}$$

Further, as always with static problems, we shall have to set

$$g_{14} = g_{24} = g_{34} = 0 \; . \; . \; . \quad \text{(E8)}$$

To calculate the remaining $g_{\mu\nu}$, Einstein assumes that a uniform distribution of mass in a finite world implies a spherical space:

It remains now to determine those components of the gravitational potential which define the purely spatial-geometrical relations of our continuum ($g_{11}, g_{12} \dots g_{33}$). From our assumption as to the uniformity of distribution of the masses generating the field, it follows that the curvature of the required space must be constant. With this distribution of mass, therefore, the required finite continuum of the $x_1$, $x_2$, $x_3$ with constant $x_4$ will be a spherical space.[23]

Einstein then derives an expression for the line element of the spherical space:

We arrive at such a space, for example, in the following way. We start from a Euclidean space of four dimensions, $\xi_1, \xi_2, \xi_3, \xi_4$, with a linear element $d\sigma$ ; let, therefore,

$$d\sigma^2 = d\xi_1^2 + d\xi_2^2 + d\xi_3^2 + d\xi_4^2 \; \dots \; . \text{(E9)}$$

In this space we consider the hyper-surface

$$R^2 = \xi_1^2 + \xi_2^2 + \xi_3^2 + \xi_4^2. \qquad \text{(E10)}$$

where *R* denotes a constant. The points of this hyper-surface form a three-dimensional continuum, a spherical space of radius of curvature *R*.

He selects only the portion of the four-dimensional hyper-surface corresponding to physical space and obtains an expression for the spatial line element by substituting for the fourth co-ordinate:

The four-dimensional Euclidean space with which we started serves only for a convenient definition of our hyper-surface. Only those points of the hyper-surface are of interest to us which have metrical properties in agreement with those of physical space with a uniform distribution of matter. For the description of this three-dimensional continuum we may employ the co-ordinates $\xi_1, \xi_2, \xi_3$ (the projection upon the hyper-plane $\xi_4 = 0$), since, by reason of (10), $\xi_4$ can be expressed in terms of $\xi_1, \xi_2, \xi_3$. Eliminating $\xi_4$ from (9), we obtain for the linear element of the spherical space the expression

$$\left.\begin{aligned} d\sigma^2 &= \gamma_{\mu\nu} d\xi_\mu d\xi_\nu \\ \gamma_{\mu\nu} &= \delta_{\mu\nu} + \frac{\xi_\mu \xi_\nu}{R^2 - \rho^2} \end{aligned}\right\} \; . \quad . \quad . \quad . \qquad \text{(E11)}$$

where $\delta_{\mu\nu} = 1$, if $\mu = \nu$; $\delta_{\mu\nu} = 0$, if $\mu \neq \nu$, and $\rho^2 = \xi_1^2 + \xi_2^2 + \xi_3^2$. The co-ordinates chosen are convenient when it is a question of examining the environment of one of the two points $\xi_1 = \xi_2 = \xi_3 = 0$.

[23] It was later pointed out that elliptical geometry was also a possibility (see section 4).

This gives him an expression for the remaining gravitational potentials:

> Now the linear element of the required four-dimensional space-time universe is also given us. For the potential $g_{\mu\nu}$, both indices of which differ from 4, we have to set
>
> $$g_{\mu\nu} = -\left(\delta_{\mu\nu} + \frac{x_\mu x_\nu}{R^2 - (x_1^2 + x_2^2 + x_3^2)}\right) \quad \text{(E12)}$$
>
> which equation, in combination with (7) and (8), perfectly defines the behaviour of measuring-rods, clocks, and light-rays in the four-dimensional world under consideration".24

## §4. On an additional term for the field equations of gravitation

Einstein now turns to the field equations of relativity, and notes that the equations are not satisfied for the energy-momentum tensor and gravitational potentials he has derived:

> My proposed field equations of gravitation for any chosen system of co-ordinates run as follows:-
>
> $$\left.\begin{aligned} G_{\mu\nu} &= -\kappa\left(T_{\mu\nu} - \frac{1}{2}g_{\mu\nu}T\right), \\ G_{\mu\nu} &= -\frac{\partial}{\partial x_\alpha}\{\mu\nu,\alpha\} + \{\mu\alpha,\beta\}\{\nu\beta,\alpha\} + \frac{\partial^2 \log\sqrt{-g}}{\partial x_\mu \partial x_\nu} - \{\mu\nu,\alpha\}\frac{\partial \log\sqrt{-g}}{\partial x_\alpha} \end{aligned}\right\} \quad \text{(E13)}$$
>
> The system of equations (13) is by no means satisfied when we insert for the $g_{\mu\nu}$ the values given in (7), (8) and (12), and for the (contravariant) energy-tensor of matter the values indicated in (6). It will be shown in the next paragraph how this calculation may conveniently be made.

Indeed, Einstein notes that his cosmic model would not be compatible with the general theory of relativity if equations (E13) were the only possible form of the field equations:

> So that, if it were certain that the field equations (13) which I have hitherto employed were the only ones compatible with the postulate of general relativity, we should probably have to conclude that the theory of relativity does not admit the hypothesis of a spatially closed universe.

The good news is then announced. An alternate formulation of the field equations exists that is both covariant and compatible with Einstein's cosmology. This formulation necessitates the introduction an extra term to the field equations, a modification that Einstein claims is *"perfectly analogous"* to that mooted in section §1 for the case of Newtonian cosmology:[25]

> However, the system of equations (13) allows a readily suggested extension which is compatible with the relativity postulate, and is perfectly analogous to the extension of Poisson's equation given by equation (2). For on the left-hand side of field equation (13) we may add the fundamental tensor $g_{\mu\nu}$ , multiplied by a universal constant, $-\lambda$,

[24] The words "in the four-dimensional world under consideration" are missing in the Perret-Jeffery translation.
[25] In fact, the modification is not "perfectly analogous", as will be discussed in section 4.

, at present unknown, without destroying the general covariance. In place of field equation (13) we write

$$G_{\mu\nu} - \lambda g_{\mu\nu} = -\kappa\left(T_{\mu\nu} - \frac{1}{2} g_{\mu\nu} T\right) \quad \dots \text{ (E13a)}$$

Einstein immediately points out that the new term will not affect relativity's successful prediction of the motion of the planets, provided the constant λ is sufficiently small. He also notes that the new formulation of the field equations is compatible with the conservation of momentum and energy:

> This field equation, with λ sufficiently small, is in any case also compatible with the facts of experience derived from the solar system. It also satisfies laws of conservation of momentum and energy, because we arrive at (13a) in place of (13) by introducing into Hamilton's principle, instead of the scalar of Riemann's tensor, this scalar increased by a universal constant; and Hamilton's principle, of course, guarantees the validity of laws of conservation.[26] It will be shown in §5 that field equation (13a) is compatible with our conjectures on field and matter.

## §5. Calculation and result

Einstein now inserts his components of the energy-momentum tensor and the $g_{\mu\nu}$ into the modified field equations. Taking the simplest case of the point (0,0,0,0), he derives two equations relating the density of matter, the radius of the cosmos and new cosmological constant, and combines them into a single equation:

> Since all points of our continuum are on an equal footing, it is sufficient to carry through the calculation for *one* point, e.g. for one of the two points with the co-ordinates
>
> $$x_1 = x_2 = x_3 = x_4 = 0.$$
>
> Then for the $g_{\mu\nu}$ in (13a) we have to insert the values
>
> $$\begin{matrix} -1 & 0 & 0 & 0 \\ 0 & -1 & 0 & 0 \\ 0 & 0 & -1 & 0 \\ 0 & 0 & 0 & +1 \end{matrix}$$
>
> wherever they appear differentiated only once or not at all. We thus obtain in the first place
>
> $$G_{\mu\nu} = -\frac{\partial}{\partial x_1}[\mu\nu, 1] + \frac{\partial}{\partial x_2}[\mu\nu, 2] + \frac{\partial}{\partial x_3}[\mu\nu, 3] + \frac{\partial^2 \log\sqrt{-g}}{\partial x_\mu \partial x_\nu}.$$
>
> From this we readily discover, taking (7), (8), and (13) into account, that all equations (13a) are satisfied if the two relations
>
> $$-\frac{2}{R^2} + \lambda = -\frac{\kappa\rho}{2}, \quad -\lambda = -\frac{\kappa\rho}{2},$$
>
> or
>
> $$\lambda = \frac{\kappa\rho}{2} = \frac{1}{R^2}. \quad . \quad . \quad \text{(E14)}$$
>
> are fulfilled.

[26] This statement draws on section 15 of the 'Grundlage' paper (Einstein 1916a).

Einstein notes that the density of matter and the radius of cosmic space are determined by the newly introduced universal λ. The mass of the universe can also be expressed:

> Thus the newly introduced universal constant λ defines both the mean density of distribution $\rho$ which can remain in equilibrium and also the radius R and the volume $2\pi^2 R^3$ of spherical space. The total mass M of the universe, according to our view, is finite, and is, in fact
>
> $$M = \rho . 2\pi^2 R^3 = 4\pi^2 \frac{R}{\kappa} = \pi^2 \sqrt{\frac{32}{\kappa^3 \rho}} . \qquad . \qquad \text{(E15)}$$

We note that Einstein makes no attempt to calculate the cosmic radius $R$ or the mass of the universe $M$ from astronomical estimates of the matter density $\rho$, as discussed in section 4. Instead, he summarizes his memoir, emphasizing that the work is merely a hypothetical model of the universe that is consistent with relativity. Whether it is compatible with observation is a question he is apparently not yet willing to address:

> Thus the theoretical view of the actual universe, if it is in correspondence with our reasoning, is the following. The curvature of space is variable in time and space, according to the distribution of matter, but we may roughly approximate to it by means of a spherical space. At any rate, this view is logically consistent, and from the standpoint of the general theory of relativity lies nearest at hand; whether, from the standpoint of present astronomical knowledge, it is tenable, will not here be discussed.

Finally, Einstein recalls that the model necessitated an extension to the field equations. He stresses that the new term is not necessitated by the postulate of positive spatial curvature, but by the postulate of a uniform distribution of matter that is approximately static:

> In order to arrive at this consistent view, we admittedly had to introduce an extension of the field equations of gravitation which is not justified by our actual knowledge of gravitation. It is to be emphasized, however, that a positive curvature of space is given by the presence of matter,[27] even if the supplementary term is not introduced. That term is necessary only for the purpose of making possible a quasi-static distribution of matter, as required by the fact of the small velocities of the stars.

## 4. Discussion of Einstein's paper

### *(i) On Einstein's view of the Newtonian universe*

On a first reading, a surprising feature of Einstein's 1917 cosmological memoir is the sizeable portion of the paper concerned with Newtonian cosmology. This analysis had two important aims. In the first instance, Einstein was no doubt pleased to show that his new theory of gravitation could not only provide a consistent model of the known universe, but could overcome a well-known puzzle associated with Newtonian cosmology. Second, an *ad-hoc* modification of Newtonian gravity provided a useful analogy for a necessary modification of the field equations of relativity.

[27] The words "the presence of matter" *("befindliche Materie")* are replaced by "the results" in the Perrett-Jeffrey translation.

In characteristic fashion, Einstein was keenly aware of a deep conundrum associated with Newtonian cosmology – its inability to account for an infinite space filled with a finite distribution of matter - but unaware of attempts to address the problem. Thus his memoir opens with the memorable sentence: *"It is well known that Poisson's equation…..is not as yet a perfect substitute for Newton's theory of action at a distance"*, but the analysis to follow proceeds from first principles. As pointed out in section 3, it is clear from Einstein's papers and correspondence of 1916 and 1917 that he was unaware of the work of Neumann or Seeliger when writing his cosmological memoir.

Einstein's assault on Newtonian cosmology in section §1 of his memoir is two-pronged. First he establishes from symmetry principles that Newtonian gravity only allows for a finite island of stars in infinite space. Then he suggests from a consideration of statistical mechanics that such an island would evaporate, in contradiction with the presumed static nature of the universe. His solution to the paradox is the introduction of a new term to Poisson's equation. This solution is very similar to that of Seeliger and Neumann, but Einstein does not appear to take it too seriously, seeing it merely as a foil for a similar modification of relativity: "*We will indicate a method that does not in itself claim to be taken seriously; it merely serves as a foil for what is to follow".* A year later, Einstein presented a simpler argument against the Newtonian universe in terms of lines of force; this argument was published in the third edition of his popular book on relativity (Einstein 1918b p123) and retained in all later editions of the book.

A few years after the publication of the 1917 memoir, the Austrian physicist Franz Selety noted that the hierarchic cosmology proposed by Carl Charlier (see section 2(ii)) avoided the paradox identified by Einstein (Selety 1922). Einstein conceded the point, but objected to the Charlier's model on the grounds that it was anti-Machian (Einstein 1922b; Einstein 1922c). Selety contested this verdict (Selety 1923; Selety 1924), but Einstein wrote no further on the subject and the hierarchic universe later fell from favour for empirical reasons.[28]

[28] See (Norton 1999; Jung 2005) for a discussion of the Einstein-Selety debate.

### *(ii) On the basic assumptions of Einstein's model*

It is clear from sections §2 and §3 of Einstein's 1917 memoir that the starting point of his cosmic model was the assumption of a universe with a static distribution of matter, uniformly distributed over the largest scales and of non-zero average density. Considering the issue of stasis first, it is generally agreed amongst historians and physicists that this assumption was entirely reasonable at the time (North 1965 pp 70-72; Ellis 1986; Kragh 2007 pp 131-132; Nussbaumer and Bieri 2009 pp 72-76; Hoefer 1994). There is no evidence that Einstein was aware of Slipher's observations of the redshift of light from the spiral nebulae, while the extra-galactic nature of the spirals had yet to be established. Indeed, many years were to elapse before the discovery of a linear relation between the recession of the distant galaxies and their distance (Hubble 1929), the first evidence for a non-static universe. Thus Einstein's assumption that *"there is a system of reference relative to which matter may be looked upon as being permanently at rest"* seems reasonable. It could perhaps be argued that Einstein erred philosophically in inferring *global* stasis from astronomical observations of the *local* environment (see for example Kerzberg 1989a; Smeenk 2014 p241). However, we find his assumption reasonable in the context of the widespread contemporaneous belief that the universe was not much larger than the Milky Way (see section (2(ii)).

It is sometimes stated that Einstein's assumption of stasis prevented him from predicting the expansion of the universe many years before the phenomenon was discovered by astronomers (see for example Fölsing 1997 p389; Isaacson 2007 p355; Ohanian 2008 p 251; Bartusiak 2009 p255). This statement may be true in a literal sense, but we find it somewhat anachronistic and in conflict with Einstein's philosophical approach to cosmology. It is clear throughout his cosmological memoir that Einstein's interest lay in establishing whether the general theory of relativity could give a consistent description of the known universe, a pragmatic approach to cosmology that was to continue in the years to follow (O'Raifeartaigh and McCann 2014). Thus, the exploration of solutions to the field equations for the case of a non-static cosmos would have been of little interest to Einstein in 1917, as discussed further in section 5. Many years later, Einstein stated that the assumption of a static universe *"appeared unavoidable to me at the time, since I thought that one would get into bottomless speculations if one departed from it"* (Einstein 1945 p137). Indeed, it could be argued that the common moniker 'Einstein's static model of the universe' is a little misleading, as it implies a choice from a smorgasbord of possible models of the known universe. Historically speaking, a more accurate title would be 'Einstein's model of the Static Universe'.

In some ways, Einstein's assumption of matter *"as being uniformly distributed over enormous spaces"* was more radical than his assumption of stasis. Technically speaking, this assumption implied a universe that was both isotropic and homogeneous, at least on the largest scales. As pointed out in section §3 of his cosmological memoir, Einstein was keenly aware that this assumption was at odds with astronomical observation of the local environment. Thus, the assumption was more of an assumed principle and indeed it was later named the 'Cosmological Principle' (Milne 1935 p24). One reason for the principle was undoubtedly simplicity, as the assumption of homogeneity and isotropy greatly simplified the business of solving the field equations. A deeper reason may have been that the Cosmological Principle chimed with a Copernican approach to cosmology and with the spirit of relativity (Bondi 1952 pp 11-13). After all, to assume a universe with a non-uniform distribution of matter on the largest scales was to assume a universe in which all viewpoints were not equivalent, in contradiction with basic tenets of relativity; indeed, we note that the Cosmological Principle was originally named 'the extended principle of relativity" (Milne 1933).

*(iii) On spatial curvature*

As described in section §2 of his memoir, Einstein's hypothesis of spherical spatial geometry arose from a consideration of the problem of boundary conditions at infinity, assuming a Machian universe with a static distribution of matter of non-zero average density. Having exhausted all other possibilities, his solution was to banish the boundaries with the postulate of closed spatial curvature. In this manner, Einstein's model of the cosmos explicitly incorporated his view of the relativity of inertia. It was later shown that closed geometry was the *only* possibility for a universe with a static, homogeneous distribution of matter of non-zero average density. Thus, Einstein's view of Mach's principle was a useful, but not strictly necessary, guide to his first model of the universe, just as it was a guide on his path to the field equations.

We recall from section 2(ii) that the notion of a universe of closed spatial geometry had been considered by some mathematicians as a theoretical possibility long before the advent of the general theory of relativity; indeed Karl Schwarzschild had noted that the phenomenon could not be ruled out on the basis of astronomical observation. It is intriguing to note that, soon after the publication of the general theory of relativity, Schwarzschild raised the issue of closed geometry for the universe in a letter to Einstein: *"As far as very large spaces are concerned, your theory takes an entirely similar position to Riemann's geometry, and you are* *certainly not unaware that elliptic geometry is derivable from your theory, if one has the entire universe under uniform pressure"* (Schwarzschild 1916b). However, the suggestion constitutes one throwaway line in a long letter and there is no indication in Einstein's response that he paid attention (Einstein 1916j).

From an aesthetic point of view, it may have been philosophically pleasing to Einstein that the structure of the universe on the largest scales should resemble the familiar form of the planets. From a geometrical point of view, the model can be described as a three-dimensional sphere embedded in four-dimensional Euclidean space. The model is often referred to as a cylindrical world, i.e., a universe divided into finite three-dimensional spatial sections distributed over cosmic time. It is sometimes stated that the model was not truly relativistic, given the independence of the time co-ordinate; Einstein was quick to refute this suggestion, noting that there was no violation of the relativity postulate for the case of a static universe (Einstein 1917f; Einstein 1922b).[29]

Following the publication of the 1917 memoir, colleagues such as Erwin Freundlich, Felix Klein and Willem de Sitter suggested to Einstein that elliptical spatial geometry was a more general possibility for his model (Einstein 1917j; Einstein 1917k; Einstein 1917l). Einstein quickly conceded the point, noting that his relation between the radius of curvature and the mean density of matter remained unchanged. For example, he remarked to Klein: *"As I have never done non-Euclidean geometry, the more obvious elliptical geometry had escaped me….my observations are just altered thus, that the space is half as large; the relation between R (the radius of curvature) and ρ (mean density of matter) is retained"* (Einstein 1917k). A few months later, he commented to de Sitter: *"When I was writing the paper, I did not yet know about the elliptical possibility…..this possibility seems more likely to me as well"* (Einstein 1917l).[30]

### *(iv) On the cosmological constant*

As described in section §4 of his cosmological memoir, Einstein soon found that the hypothesis of closed spatial geometry was not sufficient to achieve a successful relativistic model of the universe. A consistent solution could only be achieved with the introduction of an additional term $\lambda g_{\mu\nu}$ to the field equations, where $\lambda$ represented a constant that later became known as the 'cosmological constant'. Thus Einstein's model appears to have evolved according to the following sequence of assumptions: uniform, static distribution of matter → closed spatial geometry → introduction of additional term to the field equations. While the general theory allowed such a modification of the field equations, Einstein seems to have anticipated some resistance to the term; it is interesting that he forewarns the reader of what is to come on three separate occasions in the paper (in the introductory section, in his discussion of Newtonian cosmology in section §1 and in his discussion of the problem of boundary conditions in section §2). Indeed, much of Einstein's 1917 memoir can be read as a lengthy justification for the introduction of the cosmological constant term to relativity!

[29] See (Hoefer 1994) for further discussion of this point.

[30] Einstein's acceptance of elliptical geometry for his model is first mentioned in the literature in (de Sitter 1917a).

Some historians have found Einstein's use of the cosmological constant in his 1917 memoir somewhat ambiguous and argue that his view of the term wavers throughout the paper (see for example Kerzberg 1989a; Kerzberg 1989b p163). In our view, the *purpose* of the term is clear throughout the paper, both in the stated text and in the underlying physics of the model, and is summarized quite precisely in the final sentence: *"That term is necessary only for the purpose of making possible a quasi-static distribution of matter, as required by the fact of the small velocities of the stars"*. The purpose of the term was also stated clearly by Einstein in many of his reviews of the model. For example, in a 1933 review of cosmology, Einstein noted: "*Equations (1) do not allow the possibility of a non-zero uniform density of matter …. I initially found the following way out of this difficulty. The requirements of relativity permit and suggest the addition of a term of the form $\lambda g_{\mu\nu}$ to the left hand side of (1), where $\lambda$ denotes a universal constant (cosmological constant), which must be small enough that the additional term need not be considered in practice when calculating the sun's gravitational field and the motion of the planets"* (Einstein 1933). Indeed, in the same review, Einstein demonstrated from first principles that proceeding without the cosmological constant term led to the inconsistent solution $1/R^2 = 0;\ 3c^2/R^2 = \kappa\rho c^2$ (Einstein 1933).

However, although the cosmological constant term had a clear *purpose* in Einstein's memoir, there is little doubt that the term posed a significant challenge to him in terms of *interpretation.* While it is stated in section §4 of the paper that relativity *allows* the introduction of the cosmic constant term, no interpretation of the physics underlying the term is presented. Indeed, there is ample evidence that Einstein viewed the modification of the field equations as an uncomfortable mathematical necessity. Soon after the publication of the memoir, he remarked to Felix Klein: "*The new version of the theory means, formally, a complication of the foundations and will probably be looked upon by almost all our colleagues as an interesting,* *though mischievous and superfluous stunt, particularly since it is unlikely that empirical support will be obtainable in the foreseeable future. But I see the matter as a necessary addition, without which neither inertia nor geometry are truly relative*" (Einstein 1917k). More famously, in a paper of 1919, Einstein declared: *"But this view of the universe necessitated an extension of equations (1), with the introduction of a new universal constant standing in a fixed relation to the total mass of the universe (or to the equilibrium density of matter). This is gravely detrimental to the formal beauty of the theory"* (Einstein 1919b). Perhaps the best insight into Einstein's view of the term at this time can be found in a rather prescient comment to de Sitter: *"In any case, one thing stands. The general theory of relativity allows the addition of the term $\lambda g_{\mu\nu}$ in the field equations. One day, our actual knowledge of the composition of the fixed-star sky, the apparent motions of fixed stars, and the position of spectral lines as a function of distance, will probably have come far enough for us to be able to decide empirically the question of whether or not $\lambda$ vanishes. Conviction is a good mainspring, but a bad judge!"* (Einstein 1917m).

In March 1918, the Austrian physicist Erwin Schrödinger suggested that a consistent model of a static, matter-filled cosmos could be obtained from Einstein's field equations without the introduction of the cosmological constant term (Schrödinger 1918). Essentially, Schrödinger's proposal was that Einstein's solution could be obtained from the unmodified field equations (E13) if a negative-pressure term was added to the 'source' tensor on the right-hand side of the equations, i.e., by replacing Einstein's energy-momentum tensor (E6) by the tensor

$$T^{\mu\nu} = \begin{pmatrix} -p & 0 & 0 & 0 \\ 0 & -p & 0 & 0 \\ 0 & 0 & -p & 0 \\ 0 & 0 & 0 & \rho - p \end{pmatrix} \qquad (9)$$

where $\rho$ is the mean density of matter and $p$ is the pressure (defined as $p = \lambda/\kappa$ ).

Einstein's response was that Schrödinger's formulation was entirely equivalent to that of his 1917 memoir, provided the negative-pressure term was constant (Einstein 1918c).[31] This response seems at first surprising; Schrödinger's new term may have been *mathematically* equivalent to that of Einstein's but the underlying physics was surely different. However, in the same paper, Einstein gave his first physical interpretation of the cosmological term, namely that of a negative mass density: *"In terms of the Newtonian theory…a modification of the theory is required such that "empty space" takes the role of gravitating negative masses which are distributed all over the interstellar space"* (Einstein 1918c).

[31]Schrödinger also suggested that the pressure term might be time variant, anticipating the modern concept of quintessence (Schrödinger 1918), but this suggestion was too speculative for Einstein (Einstein 1918c). See (Harvey 2012) for a discussion of this episode.

Within a year, Einstein proposed a slightly different interpretation of the cosmic constant term. Rewriting the field equations in a slightly different format, he opined that the cosmological constant now took the form of a constant of integration, rather than a universal constant associated with cosmology: *"But the new formulation has this great advantage, that the quantity appears in the fundamental equations as a constant of integration, and no longer as a universal constant peculiar to the fundamental law"* (Einstein 1919b).[32] Indeed, a letter to Michele Besso suggests that Einstein had arrived at a similar interpretation a year earlier using a variational principle (Einstein 1918d). A follow-up letter to Besso suggests that at one point, Einstein considered the two views to be equivalent: *"Since the world exists as a single specimen, it is essentially the same whether a constant is given the form of one belonging to the natural laws or the form of an 'integration constant'"* (Einstein 1918e).

Thus, there is little doubt that a satisfactory interpretation of the physics associated with the cosmological constant term posed a challenge for Einstein in these years. A startling explanation for this ambiguity may be a slight confusion concerning the manner in which the term was introduced in the 1917 memoir. It is an intriguing but little-known fact that, despite his claim to the contrary, Einstein's modification of the field equations in section §4 of his memoir was not in fact *"perfectly analogous"* to his modification of Newtonian gravity in section §1. As later pointed out by several analysts,[33] the modified field equations (E13a) do not reduce in the Newtonian limit to the modified Poisson's equation (E2), but to a different relation given by

$$\nabla^2\phi + c^2\lambda = 4\pi G\rho \qquad (10)$$

This might seem a rather pedantic point, given that the general theory allowed the introduction of the cosmological constant term, irrespective of comparisons with Newtonian cosmology. Indeed, as noted in section 4(i), Einstein described his modification of Newtonian cosmology merely as a "*foil for what is to follow"*. However, the error may be significant with regard to Einstein's *interpretation* of the term. Where he intended to introduce a term to the field equations representing an attenuation of the gravitational interaction at large distances, he in fact introduced a term representing a very different effect. Indeed, the later interpretation of the cosmological term as representing a tendency for empty space to expand would have been deeply problematic for Einstein in 1917, given his understanding of Mach's Principle at the time. Thus, while there is no question that relativity allowed the introduction of the cosmic constant term, it appears that Einstein's interpretation of the term may have been to some extent founded on a misconception (Rindler 1969 p223; Harvey and Schucking 1999).

[32]In this paper, Einstein investigated whether gravitational fields play a role in the structure of elementary particles.
[33] See for example (Rindler 1969 p223; Norton 1999; Harvey and Schucking 1999; Earman 2001; Straumann 2002).

### *(v) On testing the model against observation*

A curious aspect of Einstein's 1917 memoir is that, having established a pleasing relation between the geometry of the universe and the matter it contained, he made no attempt to test the model against empirical observation. After all, even a rough estimate of the mean density of matter $\rho$ in equation (E14) would have given a value for the cosmic radius $R$, and a value for the cosmological constant $\lambda$. These values could then have been checked against observation; one could expect an estimate for $R$ that was not smaller than astronomical estimates of the size of the distance to the furthest stars, and an estimate for λ that was not too large to be compatible with observations of the orbits of the planets. No such calculation is to be found in the 1917 memoir. Instead, Einstein merely declares at the end of the paper that the model is logically consistent: *"At any rate, this view is logically consistent, and from the standpoint of the general theory of relativity lies nearest at hand; whether, from the standpoint of present astronomical knowledge, it is tenable, will not here be discussed*".

We were therefore intrigued to learn that an estimate of cosmic radius can be found in Einstein's correspondence around this time.[34] Taking a value of $\rho = 10^{-22}$ g/cm$^3$ for the mean density of matter, he obtained from equation (E14) an estimate of $10^7$ light-years for the radius of his universe. This calculation, never formally published in the literature, can be found in letters written by Einstein to Paul Ehrenfest, Erwin Freundlich, Michele Besso and Willem de Sitter in February and March 1917 (Einstein 1917c; Einstein 1917j; Einstein 1917d; Einstein 1917f).[35] In each case, Einstein appears to consider the resulting estimate of cosmic radius much too large in comparison with observation. For example, in his letter to Paul Ehrenfest, Einstein states: *"From the measured stellar densities, a universe radius of the order of magnitude of $10^7$ light-years results, thus unfortunately being very large against the distances of observable stars"* (Einstein 1917c). The 'problem' is stated more specifically in a letter to de Sitter dated March 12th: *"Astronomers have found the spatial density of matter from star counts up to the nth class….at about $10^{-22}$ g/cm$^3$. From this, approximately R = $10^7$ light-years results, whereas we only see as far as $10^4$ light-years"* (Einstein 1917f). A copy of this letter is shown in figure 3; we note that Einstein suggests in the same letter that the observation of negative parallax could offer empirical evidence for a universe of closed spatial geometry. Einstein does not specify in any of his correspondence exactly why he felt that theoretical estimates of cosmic radius should not be substantially larger than astronomical estimates of the distance to the stars, but his comments indicate that, like many of his contemporaries at this time, he did not believe that the universe was significantly larger than the Milky Way (see section 2(ii)).

[34] This was first brought to our attention by the work of George Ellis (Ellis 1986).

[35] Einstein does not give a reference for his estimate of the mean density of matter in his correspondence but it is in reasonable agreement with that given by de Sitter (de Sitter 1917a).

Why did Einstein not publish his estimate of cosmic radius in the 1917 memoir? After all, he was hardly the sort of physicist to brush apparent inconsistencies under the carpet. A likely explanation is that he lacked confidence in astronomical estimates of the mean density of matter. Some support for this explanation can be found in Einstein's letter to Freundlich mentioned above: "*The star statistics question has become a burning issue to be addressed now…..The matter of great interest here is that not only R but also ρ must be individually determinable astronomically, the latter quantity at least to a very rough approximation, and then my relation between them ought to hold. Maybe the chasm between the $10^4$ and $10^7$ light years can be bridged after all. That would mean the beginning of an epoch in astronomy"* (Einstein 1917j).

In 1921, Einstein presented a series of lectures on relativity at Princeton University, the last of which concerned the topic of general relativity and cosmology.[36] Reports of this lecture suggest that Einstein viewed the average density of matter in the universe as an unknown quantity. Indeed, as shown in figure 4, the lecture was reported in the *New York Times* under the headline 'Einstein Cannot Measure Universe' with the sub-heading "With Mean Density of Matter Unknown the Problem is Impossible". Further light on Einstein's view on the matter was given in his famous lecture 'Geometry and Experience', presented to the Prussian Academy of Sciences in January 1921: *"At first it might seem possible to determine the average*

[36] See (Illy 2005 pp 203-205; Weinstein 2013) for a description of Einstein's visit to Princeton.

*density of matter by observation of that part of the universe which is accessible to our observation. This hope is illusory. The distribution of the visible stars is extremely irregular, so that we on no account may venture to set the average density of star-matter in the universe equal to, let us say, the average density in the Galaxy"* (Einstein 1921a).

In the same lecture, Einstein suggested an astronomical method of estimating the magnitude of the cosmological constant (and thus estimating the size of the Einstein World from equation (E14)). If the statistical distribution and masses of the stars in the galaxy were known, one could calculate the minimum velocity of the stars needed to avoid gravitational collapse using Newtonian mechanics. A comparison of the observed velocity of the stars with that predicted could then give an estimate of the size of the cosmic constant (Einstein 1921a). A few months later, Einstein put this idea to the test using astronomical data for globular clusters (Einstein 1921b). The attempt was not successful, but he concluded that the method might one day succeed with more precise astronomical data: "*The incompleteness of the material presently available from observation forces us from the time being to be content with this agreement in the order of magnitude. More precise results have to be based upon a better knowledge of star masses and star velocities"* (Einstein 1921b).

### *(vi) On the stability of the Einstein World*

Perhaps the strangest aspect of Einstein's 1917 memoir is his failure to consider the stability of his cosmic model. After all, equation (E14) drew a direct equation between a universal constant $\lambda$, the radius of the universe $R$, and the density of matter $\rho$. But the quantity $\rho$ represented a *mean* value for the density of matter, arising from the theoretical assumption of a uniform distribution of matter on the largest scales. In the real universe, one would expect a natural variation in this parameter from time to time, raising the question of the stability of the model against perturbations in density. In fact, it was later shown that the Einstein World is generally unstable against such perturbations: a slight increase in the density of matter (without a corresponding change in $\lambda$) would cause the universe to contract, become more dense and contract further, while a slight decrease in density would result in a runaway expansion (Eddington 1930: Eddington 1933 pp 50-54).[37] It is more than a little curious that Einstein did not consider this aspect of his model in 1917; some years later, it was a major reason for rejecting the model, as described below.

[37] Later still, it was found that there are exceptions to this behaviour (Harrison 1967; Gibbons 1987).

## 5. Reactions to the Einstein World

At first, reactions to the Einstein World were mainly confined to a few theorists. Testing the model empirically was no trivial task; in addition, astronomical tests of the general theory itself were sparse and inconclusive in these years (Crelinsten 2006 pp 113-114; 148-152; 213-231). That said, a number of scholars attempted to use data from astronomy to calculate the size of the Einstein World over the next few years. For example, Willem de Sitter estimated a value for the radius of the Einstein World using a number of methods, such as a consideration of the apparent and known diameters of certain astronomical objects, a consideration of the lack of an antipodal image of the sun, and a consideration of the mean density of matter in the centre of the galaxy (de Sitter 1917a). The latter method proceeded in a manner identical to Einstein's calculation in his correspondence (section 4 (v) above) and a similar calculation was carried out by several other astronomers. These studies resulted in estimates of cosmic radius roughly similar to Einstein's value of $R = 10^7$ light-years, but a very different estimate was provided in 1926 by the American astronomer Edwin Hubble, whose pioneering astronomical observations expanded the cosmological distance ladder significantly. Indeed, Hubble's measurements of the distance of several spiral nebulae led him to an estimate of $1.5\text{x}10^{-31}$ g/cm$^3$ for the mean density of matter, from which he estimated a value of $10^{11}$ light-years for the radius of the Einstein World (Hubble 1926).[38]

In July 1917, Willem de Sitter noted that the modified field equations allowed an alternate cosmic solution, namely the case of a universe with no matter content (de Sitter 1917a). Approximating the known universe as an empty universe, de Sitter set the energy-momentum tensor in Einstein's extended field equations (E13a) to zero according to

$$G_{\mu\nu} - \frac{1}{2} g_{\mu\nu} G + \lambda g_{\mu\nu} = 0 \tag{11}$$

and showed that these equations have the solution

$$\rho = 0; \ \lambda = \frac{3}{R^2} \tag{12}$$

[38] See (Peruzzi and Realdi 2011) for a review of attempts to estimate the size of the Einstein World.

a result he dubbed 'Solution B' to Einstein's 'Solution A' (de Sitter 1917a). In this cosmology, Einstein's matter-filled three-dimensional universe of spherical spatial geometry was replaced by an empty four-dimensional universe of closed *spacetime* geometry.[39]

It should come as no surprise that Einstein was greatly perturbed by de Sitter's alternative cosmology. Quite apart from the fact that the empty model bore little relation to the real world, the existence of a vacuum solution for the cosmos was in direct conflict with Einstein's understanding of Mach's Principle in these years (see section 2(iii)). A long debate by correspondence ensued between the two physicists concerning the relative merits of the two models (Kerzberg 1989a; Ellis 1986; Realdi and Peruzzi 2008). Eventually, Einstein made his criticisms public in a paper of 1918: *"It appears to me that one can raise a grave argument against the admissibility of this solution…..In my opinion, the general theory of relativity is a satisfying system only if it shows that the physical qualities of space are completely determined by matter alone. Therefore no $g_{\mu\nu}$- field must exist (that is no space-time continuum is possible) without matter that generates it*" (Einstein 1918f). Einstein was no doubt pleased to find a technical objection to de Sitter's model, namely that it appeared to contain a spacetime singularity: *"However, g vanishes also for $r = \frac{\pi}{2}R$, and it seems that no choice of co-ordinates can remove this discontinuity…Until the opposite is proven, we have to assume that the de Sitter solution has a genuine singularity on the surface $r = \frac{\pi}{2}R$ in the finite domain; i.e., it does not satisfy the field equations…for any choice of co-ordinates"* (Einstein 1918f). Indeed, Einstein took the view that the de Sitter universe was not truly empty, but that its matter was contained at the horizon: *"The de Sitter system does not look at all like a world free of matter, but rather like a world whose matter is concentrated entirely on the surface $r = \frac{\pi}{2}R$"* (Einstein 1918f).

In the years that followed, Einstein continued to debate the relative merits of 'Solution A' and 'Solution B' with de Sitter and other physicists such as Kornel Lanczos, Hermann Weyl, Felix Klein and Gustav Mie. Throughout this debate, Einstein did not waver from his core belief that a satisfactory cosmology should describe a universe that was globally static with a metric structure that was fully determined by matter.[40] In correspondence with Felix Klein (Klein 1918; Einstein 1918g), Einstein eventually conceded that the apparent singularity in the de Sitter universe was an artefact of co-ordinate representation: *"My critical remark about de*

[39] Speaking technically, the gravitational potentials vanish at both spatial and temporal infinity in de Sitter's model.

[40] See (Schulmann et al. 1988 pp 351-352) for a discussion of the so-called Einstein-deSitter-Weyl-Klein debate.

*Sitter's solution needs correction; a singularity-free solution for the gravitation equations without matter does in fact exist".* However, he noted in the same letter that the concession applied only for the case of a non-static universe, a solution he considered unrealistic: "*However, under no condition could this world come under consideration as a physical possibility. For in this world, time t cannot be defined in such a way that that the three-dimensional slices t = const. do not intersect one another…"* (Einstein 1918g). It is noteworthy that Einstein never formally retracted his criticism of the de Sitter universe in the literature, nor did he refer to the de Sitter model in his discussions of cosmology in his popular book on relativity (Einstein 1918b p116), his Princeton lectures (Einstein 1922d pp 110-111) or his 1921 essay on geometry and the universe (Einstein 1921a).

Despite Einstein's reservations, the de Sitter model attracted a great deal of interest amongst both theorists and astronomers. The principle reason for this was a prediction that light emitted by an object placed in the de Sitter universe would be red-shifted, a phenomenon that became known as the 'de Sitter effect'.[41] This prediction chimed with emerging observations of the spectra of the spiral nebulae (see section 2(ii)) and theorists such as Kornel Lanczos, Arthur Stanley Eddington, Hermann Weyl and Howard Percy Robertson published detailed analyses of the de Sitter model (Lanczos 1922; Weyl 1923a, 1923b; Eddington 1923; Robertson 1928, 1929).[42] Meanwhile, astronomers such as Karl Wirtz, Ludwig Silberstein, Knut Lundmark and Gustav Strömberg (Wirtz 1922; Silberstein 1924; Lundmark 1924; Strömberg 1925) sought to measure the curvature of the de Sitter universe from astronomical observations of celestial objects such as B stars, globular clusters, novae and nebulae. However, these attempts to match theory with observation were not successful due to a lack of knowledge of the true distance of many of these astronomical objects, and due to a mathematical confusion concerning the nature of the de Sitter universe (Smith 1979; Nussbaumer and Bieri 2009 pp 96-98).

In 1922, the Russian physicist Alexander Friedman suggested that non-static solutions of the Einstein field equations should be considered in relativistic models of the cosmos (Friedman 1922). Starting from the modified field equations (E13a) and assuming a positive spatial curvature for the cosmos, he derived two differential equations linking the time evolution of the cosmic radius $R$ with the comic density $\rho$ and the cosmological constant $\lambda$. Few physicists paid attention to Friedman's time-varying cosmology, possibly because the work was quite technical and made no connection to astronomy. Worse, Einstein publicly faulted Friedman's analysis on the basis that it contained a mathematical error (Einstein 1922a). When it transpired that the error lay in Einstein's criticism, it was duly retracted (Einstein 1923a). However, an unpublished draft of Einstein's retraction demonstrates that he did not consider Friedman's cosmology to be realistic: *"to this a physical significance can hardly be ascribed*" (Einstein 1923b).[43]

[41] In fact, the model predicted two distinct redshift effects (de Sitter 1917a).
[42] In retrospect, these analyses represented non-static solutions to the field equations, although this was not realised at the time.

A few years later, the Belgian physicist Georges Lemaître independently derived differential equations for the radius of the cosmos from Einstein's modified field equations (E13a). Aware of Slipher's observations of the redshifts of the spiral nebulae, and of emerging measurements of the distance of the spirals by Edwin Hubble (see section 2(ii)), Lemaître suggested that the recession of the nebulae was a manifestation of the expansion of space from a pre-existing Einstein World of cosmic radius $R_0 = 1/\sqrt{\lambda}$ (Lemaître 1927). This work also received very little attention at first, probably because it was published in a little-read Belgian journal. The work was brought to Einstein's attention by Lemaître himself, only to be dismissed as "abominable" (Lemaître 1958). According to Lemaître, Einstein's rejection probably stemmed from a lack of knowledge of developments in astronomy: *"Je parlais de vitesses des nébeleuses et j'eus l'impression que Einstein n'était guère au courant des faits astronomiques"* (Lemaître 1958).

In 1929, Edwin Hubble published the first evidence of a linear relation between the redshifts of the spiral nebulae and their radial distance (Hubble 1929). Soon, a variety of relativistic time-varying models of the cosmos were proposed (Eddington 1930, 1931: de Sitter 1930a, 1930b; Tolman 1930a, 1930b, 1931a, 1932; Heckmann 1931, 1932; Robertson 1932, 1933; Lemaître 1931a; Lemaître 1933). Few of these models considered the question of cosmic origins, but Eddington favoured a universe that expanded from an initial static Einstein World (Eddington 1930; Eddington 1931), not unlike Lemaître's model of 1927.[44]

As for Einstein, he made several public statements during a sojourn in California in 1931 to the effect that he accepted Hubble's observations as likely evidence of a non-static universe. For example, the *New York Times* reported Einstein as commenting that *"New observations by Hubble and Humason concerning the redshift of light in distant nebulae make the presumptions near that the general structure of the universe is not static*" (AP 1931a) and "*The redshift of the distant nebulae have smashed my old construction like a hammer blow"* (AP 1931b). In

[43] Einstein withdrew the remark before publication. A detailed account of this episode can be found in (Stachel 1977; Nussbaumer and Bieri 2009 pp 91-92).
[44] For this reason, Eddington's model became known as the Eddington-Lemaître model.

April 1931, Einstein published his first model of the expanding cosmos (Einstein 1931). Starting with Friedman's 1922 analysis of a matter-filled dynamic universe of positive spatial curvature, he removed the cosmological term from the field equations and derived simple expressions relating the rate of cosmic expansion (as measured from the recession of the nebulae), to the radius of the cosmos, the density of matter and the timespan of the expansion. It is interesting to note that Einstein provided a two-fold justification for abandoning the cosmic constant term in this paper. In the first instance, the term was unsatisfactory because it did not provide a stable solution: "*It can also be shown… that this solution is not stable. On these grounds alone, I am no longer inclined to ascribe a physical meaning to my former solution"* (Einstein 1931). In the second instance, the term was unnecessary because the assumption of stasis was no longer justified by observation: *"Now that it has become clear from Hubbel's* [sic] *results that the extra-galactic nebulae are uniformly distributed throughout space and are in dilatory motion (at least if their systematic redshifts are to be interpreted as Doppler effects), assumption (2) concerning the static nature of space has no longer any justification*" (Einstein 1931).[45] A year later, Einstein proposed an even simpler model of the expanding universe, once again with the cosmic constant term removed (Einstein and de Sitter 1932).

Thus it is clear that, when presented with empirical evidence for a dynamic universe, Einstein lost little time in abandoning his static cosmology.[46] He also abandoned the cosmological constant term and was never to re-instate it to the field equations, despite the reservations of colleagues.[47] Indeed, he is reputed to have described the term in later years as *"my biggest blunder".* Whether Einstein used these exact words may never be known,[48] but his considered view of the cosmological constant was made clear in a 1945 review of relativistic cosmology: *"If Hubble's expansion had been discovered at the time of the creation of the general theory of relativity, the cosmologic member would never have been introduced. It seems now so much less justified to introduce such a member into the field equations, since*

---

[45] An early portend of this strategy can be found on a postcard written by Einstein to Hermann Weyl in 1923. In response to Weyl's discussion of the de Sitter universe, Einstein wrote *"if there is no quasi-static world after all, then away with the cosmological term"* (Einstein 1923c). See also (Straumann 2002; Nussbaumer and Bieri 2009 pp 82-83).
[46] It is now known that Einstein also attempted a steady-state model of the expanding universe in early 1931 but abandoned the model before publication (O'Raifeartaigh et al 2014; Nussbaumer 2014b).
[47] Many physicists such as Richard Tolman, Arthur Stanley Eddington and Georges Lemaître felt the term served a useful function in addressing problems such as the timespan of the expansion and the formation of the galaxies (Tolman 1931b; Eddington 1933 p104; Lemaître 1933).
[48] This statement was reported by the Russian physicist George Gamow (Gamow 1956; Gamow 1970 p44). Some doubt has been cast on the accuracy of Gamow's report in recent years (Straumann 2008; Livio 2013 pp 231-243), while the report has been supported by Ralph Alpher (Topper 2013 p165) and by John Archibald Wheeler (Taylor and Wheeler 2000 pG-11).

*its introduction loses its sole original justification – that of leading to a natural solution of the cosmologic problem"* (Einstein 1945 p130). This passage provides further evidence of Einstein's pragmatic approach to cosmology. If the known universe could be modelled without the cosmic constant term, why include it? It is tempting to state that Einstein would have been wiser to leave the term undetermined; however, such a view is somewhat anachronistic as observational evidence for a non-zero cosmological constant did not emerge until the end of the century.[49]

## 6. Conclusions

We note in conclusion that the first relativistic model of the universe was firmly grounded in reality. In his 1917 cosmological memoir, Einstein demonstrated that the newly-minted general theory of relativity could give a consistent model of the known universe that accorded with his views on the relativity of inertia. The price was the hypothesis of closed spatial geometry for the cosmos and a modification of the field equations of general relativity. The Einstein World was not a favourite selection from a smorgasbord of possible models, but the only consistent relativistic model of a static universe with an average density of matter that differed from zero.

It is intriguing that a mathematical oversight may have been responsible for a slight confusion in Einstein's interpretation of the role of the cosmological constant; this fact should be better known. It is also interesting that Einstein made no attempt to test his model against empirical observation; later writings suggest that he distrusted astronomical estimates of the mean density of matter in the universe. Perhaps the most surprising aspect of Einstein's 1917 memoir is his failure to consider the stability of his cosmic model. When he finally abandoned the Einstein World in 1931, it was on the twin grounds that the model was both unstable and in conflict with empirical observation.

### *Coda: the emergent universe*

We note finally that the Einstein World has become a topic of renewed interest in today's cosmology. This may seem at first surprising, given the observational evidence for an expanding universe. However, many theorists have become interested in the hypothesis of a universe that inflates from a static Einstein World after an indefinite period of time, thus reviving the Eddington-Lemaître model in the context of the modern theory of cosmic inflation. It is thought that this scenario, known as 'the emergent universe', might avoid major difficulties in modern cosmology such as the horizon problem, the quantum gravity era and the initial singularity.[50] While we saw in section 4(vi) that the Einstein World is not generally stable against simple perturbations in density, a different scenario may apply in situations where quantized gravitational effects can be expected to be significant. Thus, it is intriguing to encounter intense research into the stability of the Einstein World in the context of contemporary theories of gravitation such as Brans-Dicke theory (Huang *et al.*. 2014), Einstein-Cartan theory (Atazadeh 2014), doubly general relativity (Khodadi *et al.* 2015), massive gravity (Parisi *et al.* 2012), *f*(R) gravity (Seahra and Bohmer 2009), *f*(RT) gravity (Shabani and Ziaie 2016) and loop quantum gravity (Parisi *et al.* 2007). Whether the emergent universe will offer a plausible, consistent description of the origins and evolution of our universe is not yet known, but we note, as so often, the relevance of past models of the universe in today's research.

[49] See (Earman 2001; Kragh and Overduin 2014 pp 101-109) for a review.

**Acknowledgements**

The authors wish to acknowledge the use of online materials provided by the Einstein Papers Project, an important historical resource published by Princeton University Press in conjunction with the California Institute of Technology and the Hebrew University of Jerusalem. We would also like to thank the Hebrew University of Jerusalem for permission to show the materials shown in figures 1-4 and to thank Dr. Brendan McCann and Dr. Cordula Weiss for advice on German-English translations. Cormac O'Raifeartaigh thanks the Dublin Institute for Advanced Studies for the use of research facilities, and Professor G.F. R. Ellis, Professor Jim Peebles and Dr. Kevin Brown for helpful discussions. Simon Mitton thanks St Edmund's College of the University of Cambridge for supporting his research in the history of science.

[50] See (Barrow *et al.* 2003; Ellis and Maartens 2004) for an introduction to emergent cosmology.

**Figure 1** The title page of Einstein's paper '*Kosmologische Betrachtungen zur allgemeinen Relativitätstheorie'* as it appeared in the journal *Sitzungsberichte der Königlich Preussischen Akademie der Wissenschaften* (Einstein 1917a).

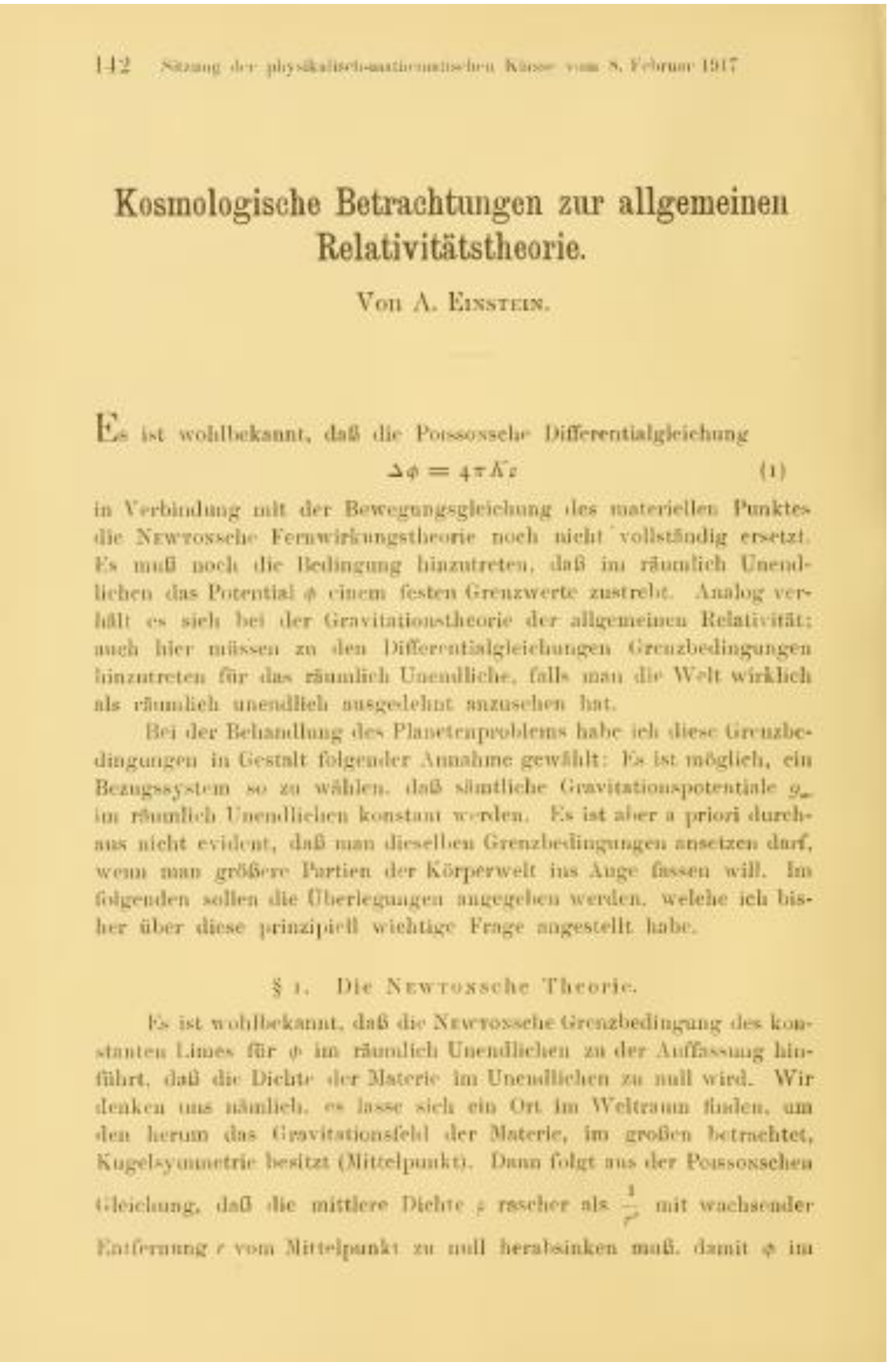

142 Sitzung der physikalisch-mathematischen Klasse vom 8. Februar 1917

# Kosmologische Betrachtungen zur allgemeinen Relativitätstheorie.

Von A. Einstein.

Es ist wohlbekannt, daß die Poissonsche Differentialgleichung

$$\Delta\phi = 4\pi K\rho \quad (1)$$

in Verbindung mit der Bewegungsgleichung des materiellen Punktes die Newtonsche Fernwirkungstheorie noch nicht vollständig ersetzt. Es muß noch die Bedingung hinzutreten, daß im räumlich Unendlichen das Potential $\phi$ einem festen Grenzwerte zustrebt. Analog verhält es sich bei der Gravitationstheorie der allgemeinen Relativität; auch hier müssen zu den Differentialgleichungen Grenzbedingungen hinzutreten für das räumlich Unendliche, falls man die Welt wirklich als räumlich unendlich ausgedehnt anzusehen hat.

Bei der Behandlung des Planetenproblems habe ich diese Grenzbedingungen in Gestalt folgender Annahme gewählt: Es ist möglich, ein Bezugssystem so zu wählen, daß sämtliche Gravitationspotentiale $g_{\mu\nu}$ im räumlich Unendlichen konstant werden. Es ist aber a priori durchaus nicht evident, daß man dieselben Grenzbedingungen ansetzen darf, wenn man größere Partien der Körperwelt ins Auge fassen will. Im folgenden sollen die Überlegungen angegeben werden, welche ich bisher über diese prinzipiell wichtige Frage angestellt habe.

## § 1. Die Newtonsche Theorie.

Es ist wohlbekannt, daß die Newtonsche Grenzbedingung des konstanten Limes für $\phi$ im räumlich Unendlichen zu der Auffassung hinführt, daß die Dichte der Materie im Unendlichen zu null wird. Wir denken uns nämlich, es lasse sich ein Ort im Weltraum finden, um den herum das Gravitationsfeld der Materie, im großen betrachtet, Kugelsymmetrie besitzt (Mittelpunkt). Dann folgt aus der Poissonschen Gleichung, daß die mittlere Dichte $\rho$ rascher als $\frac{1}{r^2}$ mit wachsender Entfernung $r$ vom Mittelpunkt zu null herabsinken muß, damit $\phi$ im

**Figure 2** The fifth page of Einstein's handwritten manuscript '*Kosmologische Betrachtungen zur allgemeinen Relativitätstheorie'*, the only surviving fragment of the original manuscript.

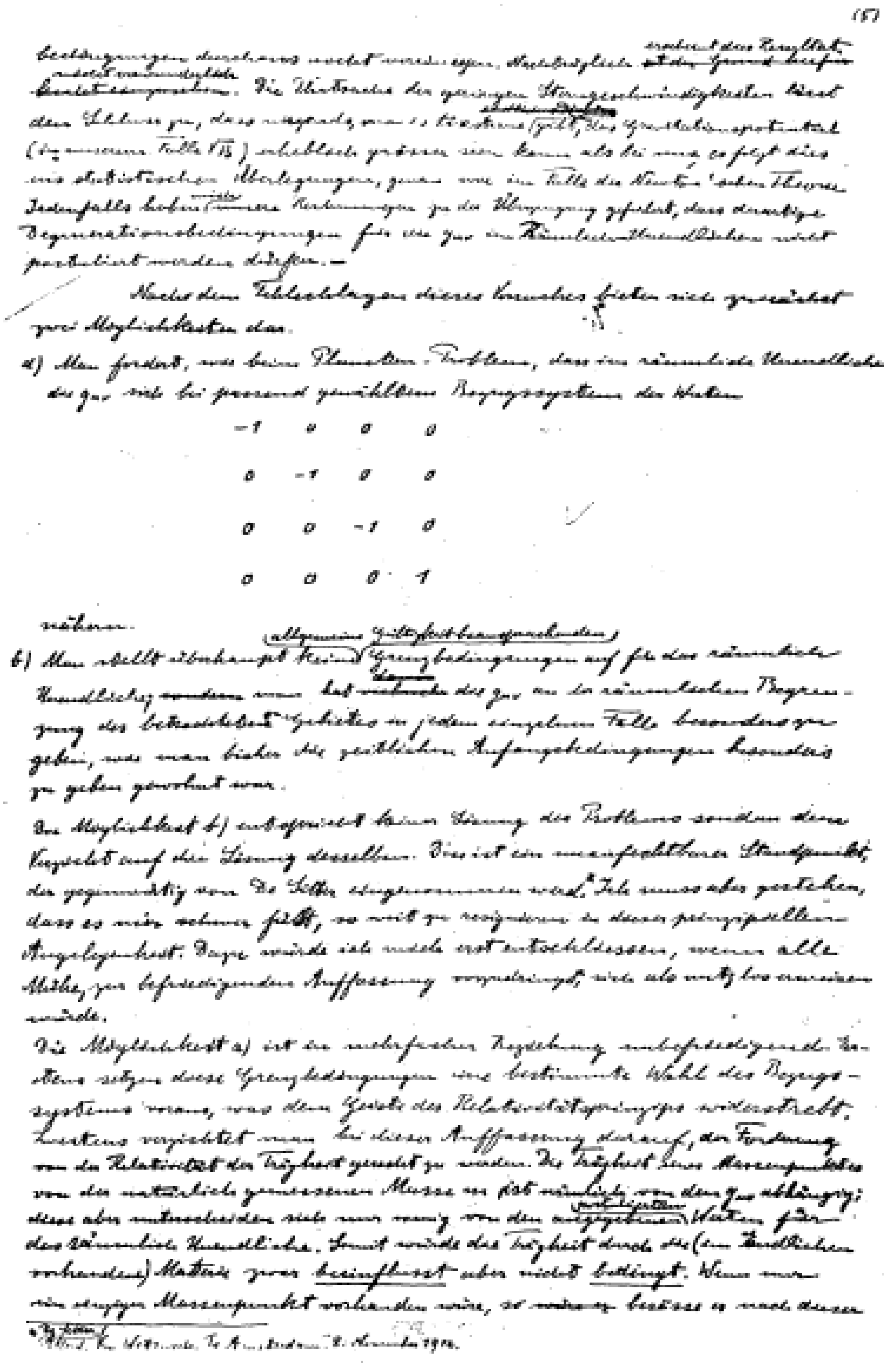

**Figure 3.** Einstein's handwritten letter to Willem de Sitter of March 12th 1917, reproduced from the Albert Einstein Archive of the Hebrew University of Jerusalem. Taking a value of $10^{-22}$ g/cm$^3$ for the mean density of matter, Einstein calculates a value of $10^7$ light-years for the radius of the cosmos and compares it with an observational value of $10^4$ light-years.

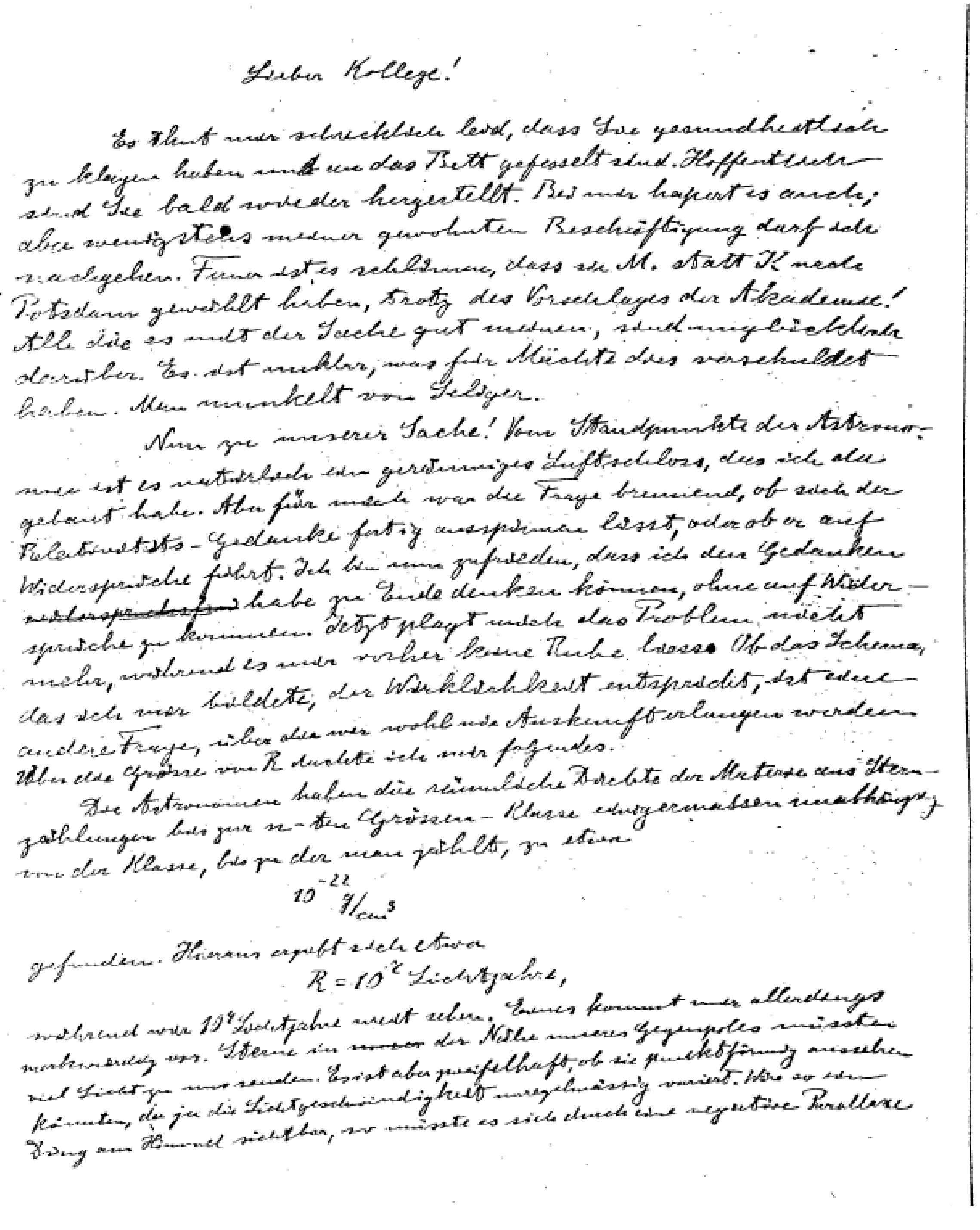

Lieber Kollege!

Es thut mir schrecklich leid, dass Sie gesundheitlich zu klagen haben und an das Bett gefesselt sind. Hoffentlich sind Sie bald wieder hergestellt. Bei mir hapert es auch; aber wenigstens meiner gewohnten Beschäftigung darf ich nachgehen. Ferner ist es schlimm, dass sie M. statt K. nach Potsdam gewählt haben, trotz des Vorschlages der Akademie! Alle die es mit der Sache gut meinen, sind unglücklich darüber. Es ist unklar, was für Mächte dies verschuldet haben. Man munkelt von Seliger.

Nun zu unserer Sache! Vom Standpunkte der Astronomie ist es natürlich ein geräumiges Luftschloss, das ich da gebaut habe. Aber für mich war die Frage brennend, ob sich der Relativitäts-Gedanke fertig ausspinnen lässt, oder ob er auf Widersprüche führt. Ich bin nun zufrieden, dass ich den Gedanken ~~widerspruchsfrei~~ habe zu Ende denken können, ohne auf Widersprüche zu kommen. Jetzt plagt mich das Problem nicht mehr, während es mir vorher keine Ruhe liess. Ob das Schema, das ich mir bildete, der Wirklichkeit entspricht, ist eine andere Frage, über die wir wohl nie Auskunft erlangen werden.

Über die Grösse von R dachte ich mir folgendes. Die Astronomen haben die räumliche Dichte der Materie aus Sternzählungen bis zur n-ten Grössen-Klasse einigermassen unabhängig von der Klasse, bis zu der man zählt, zu etwa

$$10^{-22} \, g/cm^3$$

gefunden. Hieraus ergibt sich etwa

$$R = 10^7 \text{ Lichtjahre},$$

während wir $10^4$ Lichtjahre weit sehen. Eines kommt mir allerdings merkwürdig vor. Sterne in ~~[illegible]~~ der Nähe unseres Gegenpoles müssten viel Licht zu uns senden. Es ist aber zweifelhaft, ob sie punktförmig aussehen könnten, da ja die Lichtgeschwindigkeit unregelmässig variiert. Wäre so ein Ding am Himmel sichtbar, so müsste es sich durch eine negative Parallaxe

**Figure 4.** Report in the *New York Times*, May 14th 1921, describing Einstein's lecture on relativistic cosmology at Princeton University. Acording to the report, Einstein was of the view that the size of the universe could not be estimated from his model because the mean density of matter was an unknown quantity.

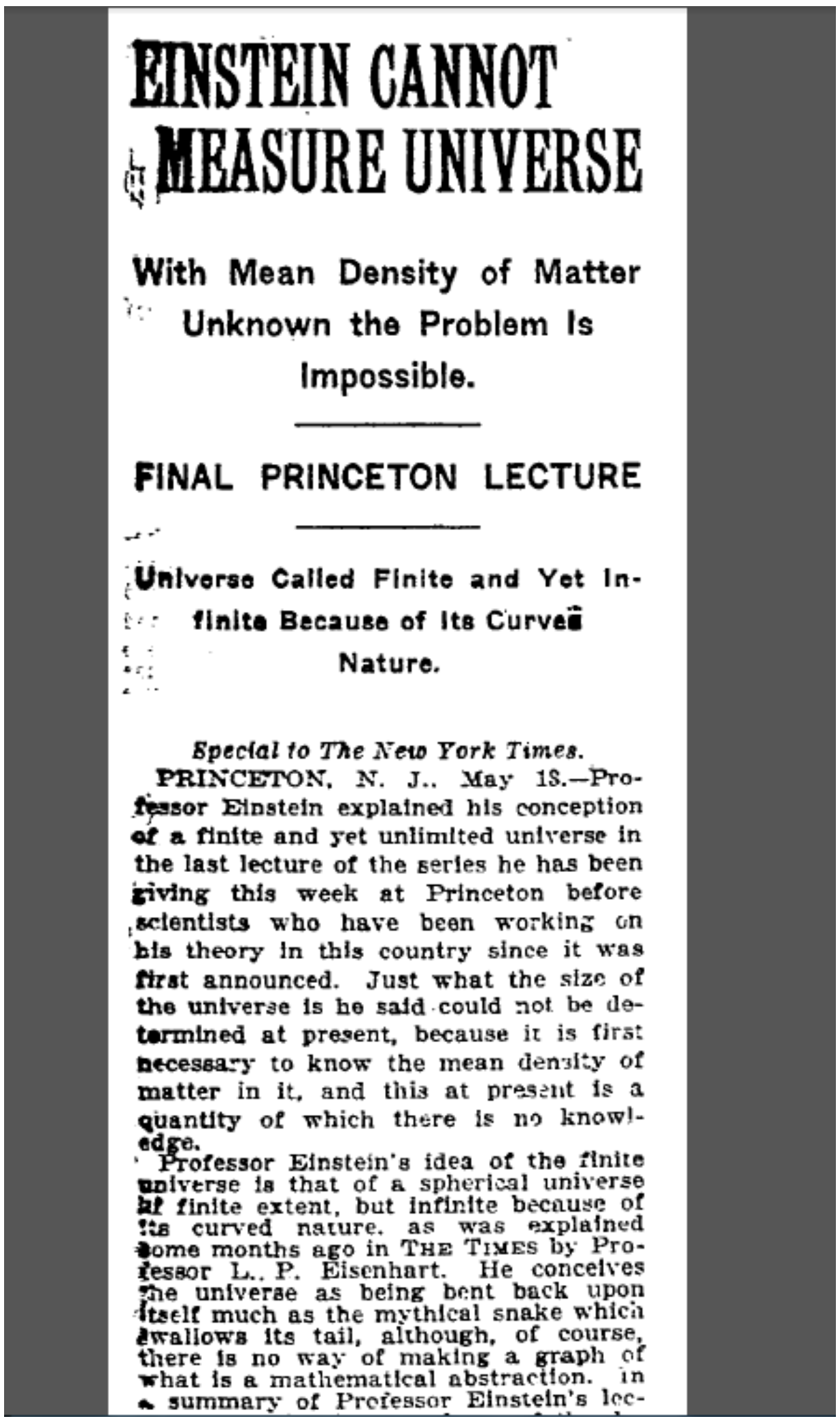

# EINSTEIN CANNOT MEASURE UNIVERSE

## With Mean Density of Matter Unknown the Problem Is Impossible.

## FINAL PRINCETON LECTURE

### Universe Called Finite and Yet Infinite Because of Its Curved Nature.

*Special to The New York Times.*

PRINCETON, N. J., May 13.—Professor Einstein explained his conception of a finite and yet unlimited universe in the last lecture of the series he has been giving this week at Princeton before scientists who have been working on his theory in this country since it was first announced. Just what the size of the universe is he said could not be determined at present, because it is first necessary to know the mean density of matter in it, and this at present is a quantity of which there is no knowledge.

Professor Einstein's idea of the finite universe is that of a spherical universe of finite extent, but infinite because of its curved nature, as was explained some months ago in THE TIMES by Professor L. P. Eisenhart. He conceives the universe as being bent back upon itself much as the mythical snake which swallows its tail, although, of course, there is no way of making a graph of what is a mathematical abstraction. In a summary of Professor Einstein's lec-